\documentclass
[preprint,10pt,psfig,nofootinbib,showpacs,nobibnotes,superscriptaddress,balancelastpage,
,titlepage,twocolumn,pre]{revtex4}%
\usepackage{amsfonts}
\usepackage{amsmath}
\usepackage{amssymb}
\usepackage{graphicx}%
\begin{document}
\preprint{ }
\title{Quasi-stationary evolution of systems driven by particle evaporation}
\author{L. Velazquez}
\email{luisberis@geo.upr.edu.cu}
\affiliation{Departamento de F\'{\i}sica, Universidad de Pinar del R\'{\i}o, Mart\'{\i}
270, Esq. 27 de Noviembre, Pinar del R\'{\i}o, Cuba.}
\author{H. Mosquera Cuesta}
\email{hermanjc@cbpf.br}
\affiliation{Centro Brasileiro de Pesquisas F\'{\i}sicas, Laborat\'{o}rio de Cosmologia e
F\'{\i}sica Experimental de Altas Energias, Rua Dr. Xavier Sigaud 150, CEP
22290-180, Urca, Rio de Janeiro, RJ, Brazil}
\author{F. Guzm\'{a}n}
\email{guzman@info.isctn.edu.cu}
\affiliation{Departamento de F\'{\i}sica Nuclear, Instituto Superior de Ciencias y
Tecnolog\'{\i}a Nucleares, Carlos III y Luaces, Plaza, La Habana, Cuba.}
\date{\today}

\begin{abstract}
We study the \textit{quasi-stationary} evolution of systems where an
\textit{energetic confinement} is unable to completely retain their
constituents. It is performed an extensive numerical study of a gas whose
dynamics is driven by binary encounters and its particles are able to
escape\ from the container when their kinetic energies overcome a given cutoff
$u_{c}$. We use a parametric family of differential cross sections in order to
modify the effectiveness of this equilibration mechanism. It is verified that
when the binary encounters favor an effective exploration of all accessible
velocities, the quasi-stationary evolution is reached when the detailed
balance is imposed for all those binary collisions which do not provoke
particle evaporation. However, the weakening of this effectiveness leads to
energy distribution functions which could be very well fitted by using a
Michie-King-like profile. We perform a theoretical analysis, in the context of
Hamiltonian systems driven by a strong chaotic dynamics and particle
evaporation, in order to take into account the effect of the nonhomogeneous
character of the confining potential.

\end{abstract}
\maketitle

\section{Introduction}

Gravitation is a generic example of a selfgravitating interaction which is not
able to effectively confine particles: it is always possible that some of them
reach the sufficient energy for escaping from a putative system, so that the
astrophysical systems undergo an evaporation process, and therefore, they will
never be in a \textit{real} thermodynamic equilibrium. This is precisely the
origin of the \textit{long-range singularity} of the thermodynamical
description of astrophysical systems
\cite{antonov,Lynden,ruelle,lind,thirring,kies,pad,gross,de vega,chava}.

On the other hand, the N-body gravitating systems satisfy a strong chaos
criterion (see Ref.\cite{pet}), supporting the assumption of an increasingly
uniform spreading of orbits over the constant energy surface or its
intersection with any other integral of motion as the angular momentum. The
competition of these dynamical behaviors produces the achievement of a
\textit{quasi-stationary regime} in the evolution of the astrophysical
systems, allowing in this way the applicability of some ordinary equilibrium
statistical mechanics methods
\cite{antonov,Lynden,ruelle,lind,thirring,kies,pad,gross,de vega,chava,pet}.

The aim of the present paper is to study the quasi-stationary evolution of
systems driven by a conservative microscopic dynamics which is also strongly
affected by particle evaporation. Our focus will be mainly on the analysis of
the effect of evaporation under the modification of the efficiency of the
equilibration mechanisms (sections \ref{efficient} and \ref{generic} ). The
interest in such study is justified as follows.

As discussed elsewhere, the equilibration mechanisms are closely connected
with the chaotic properties of the microscopic dynamics. This last ones
explain both, qualitatively and quantitatively, whether or not the dynamics
provides an effective exploration of all those accessible microscopic
configurations, and consequently, of how strong is the system tendency towards
its equilibration \cite{pet,arnold,sinai,fermi,cohen,krylov}. Hence, it is
clearly evident that the influence of a dissipative process, as the particle
evaporation, is significantly affected by how efficient are the system's
equilibration mechanisms.

On the other hand, as a by-product of the above study, we will perform in
section \ref{nonhomogeneous} a theoretical analysis of the effect of
evaporation on the distribution function over the single particle
six-dimensional $\mu$-phase space, $f\left(  \mathbf{r},\mathbf{p}\right)  $.
This analysis is carried out in the context of Hamiltonian systems driven by a
very strong chaotic dynamics, and our interest is the consideration of the
nonhomogeneous character of the confining interaction. The result is
contrasted with the requirements of the Jeans theorem \cite{jeans} for the
collisionless selfgravitating systems, as well as to the Michie-King model of
globular clusters and elliptical galaxies \cite{king,spitzer,chandra,bin}.

\section{A simple model\label{efficient}}

It was done in the past an important set of works devoted to the analysis of
particle evaporation in the context of astrophysical systems in the framework
of binary encounters (see for example in \cite{king,spitzer,chandra,bin} and
references therein). We are interested in performing a numerical closer look
at of these studies by considering the binary collisions from a perspective
which allows us to control the effectiveness of this equilibration mechanism.

Considering the general lines of the classical paper of Spitzer and H\"{a}rm
\cite{spitzer}, we assume that the following \textit{idealizations} procee:
The potential energy is \textit{constant} within the volume of the system with
spherical symmetry, but vanishes in the outside region:%

\begin{equation}
u\left(  r\right)  =\left\{
\begin{array}
[c]{cc}%
-u_{c} & r\leq R\\
0 & r>R
\end{array}
\right.  .
\end{equation}

This first idealization disregards the effects of the nonhomogeneous character
of the confining interaction, and therefore, this self-gravitating model is
equivalent to a gas of particles enclosed in a special container which allows
a given particle to escape when its kinetic energy overcomes a given cutoff
$u_{c}$. The second idealization is to assume that the particle mean-free-path
is many times the characteristic size (or length scale) of the system, which
allows us to take into account only binary collisions. Consequently, a given
particle that gains sufficient velocity after a collision will evaporate,
since the probability that this particle loses the excessive energy before
escaping from container is very small. The third idealization is to suppose
that the system is embedded in a bidimensional space.

The following \textit{toy dynamics }will be taken into account in order to
perform a numerical study of the above model. Let us consider a system with
$N$ particles and assign to each of them a vector velocity $\mathbf{v}$ and a
binary number $b$, being $\mathbf{v}_{k}$ and $b_{k}$ the corresponding data
for the \textit{k}-th particle. The binary number $b=0$ indicates that the
corresponding particle has escaped from the system, or to the contrary, that
it remains trapped when $b=1$. The dynamical evolution of the model system is
carried out as follows:

\begin{itemize}
\item For each computational step, a sample of $M$ interacting pairs are
chosen one by one at random.

\item The pair interaction is suppressed when at least one particle has
already escaped from the system.

\item When the pair interaction is possible, the particles change their
velocities as follows:%
\begin{equation}
\mathbf{\tilde{v}}_{1,2}=\frac{\mathbf{v}_{1}+\mathbf{v}_{2}}{2}\pm\left\vert
\frac{\mathbf{v}_{1}-\mathbf{v}_{2}}{2}\right\vert \mathbf{n}, \label{din1}%
\end{equation}
being $\mathbf{n}$ an unitary vector defined by:%
\begin{equation}
\mathbf{n}=\left[  \cos\left(  \theta_{0}+\theta\right)  ,\sin\left(
\theta_{0}+\theta\right)  \right]  ,
\end{equation}
where the phase $\theta$ takes a random value belonging to the interval
$\left(  -\pi,\pi\right)  $, being $\theta_{0}$ the phase of the relative
velocity $\mathbf{v}=\left(  \mathbf{v}_{1}-\mathbf{v}_{2}\right)  /2$. The
above equation ensures the energy and linear momentum conservation of the pair
during the collision. If after the collision the kinetic energy of a given
particle of the pair is greater than\ the escape energy $u_{c}$, its
corresponding binary number changes to zero, which means that this particle
has escaped from the system.

\item The random phase $\theta$ is distributed according to the following
generic differential cross section:%
\begin{equation}
d\sigma\left(  \theta;a\right)  =\frac{\sigma}{2K}\frac{ad\theta}{a^{2}%
+\theta^{2}}, \label{dCS2}%
\end{equation}
where $\sigma$ is the total cross section, $a$, a positive real parameter
which drives the form of the interaction, and $K$, a normalization constant,
$K=\arctan\left(  \pi/a\right)  $.
\end{itemize}

The specific form of the differential cross section (\ref{dCS2}) allows us to
carry out a comparative study of the quasi-stationary evolution features when
the efficiency of the equilibration mechanism is modified: the variation of
the deformation parameter $a$ affects the way that the binary encounters allow
the particles to explore the velocities which are stable under the evaporation.

\begin{figure}
[ptb]
\begin{center}
\includegraphics[
height=3.039in,
width=3.2396in
]%
{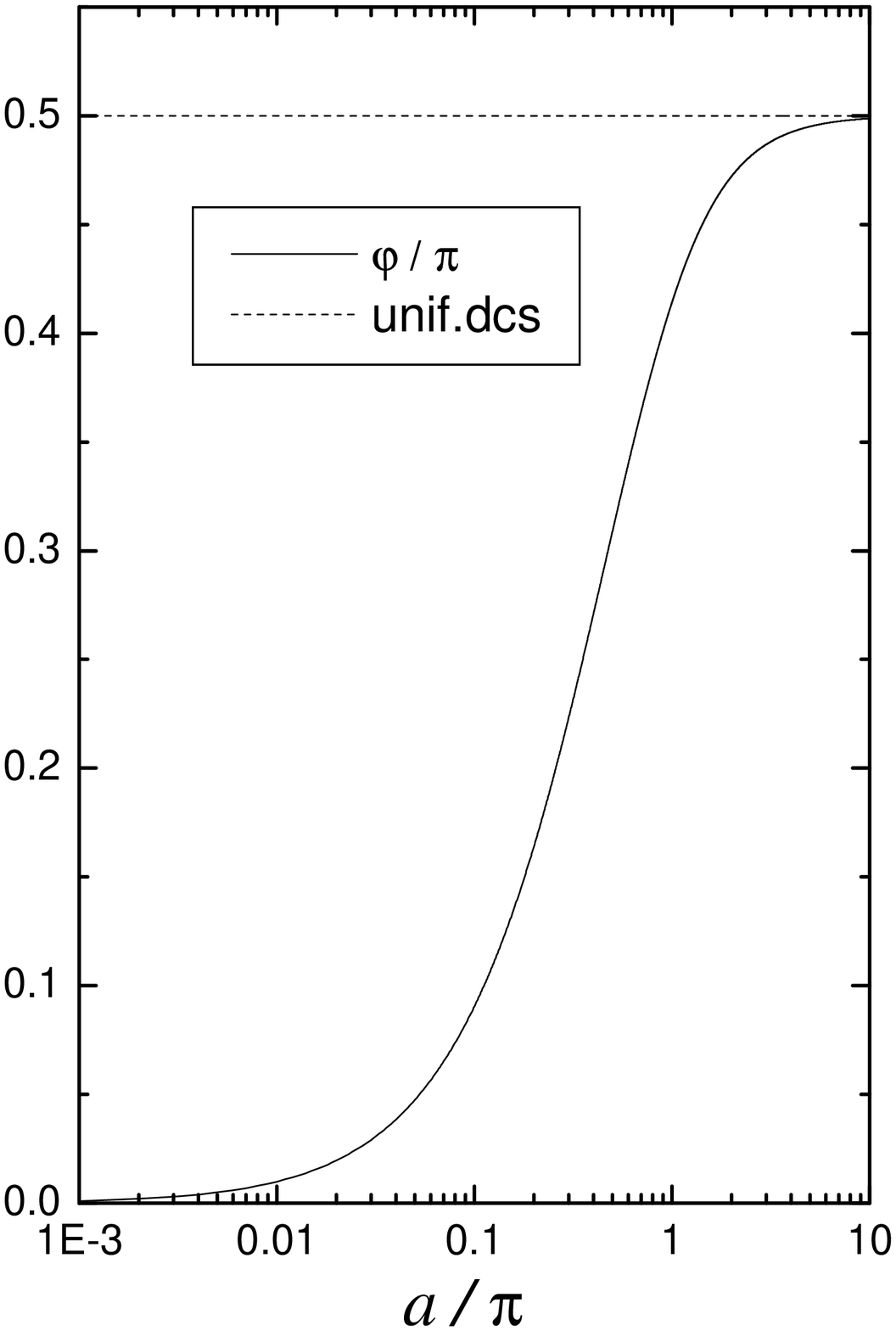}%
\caption{Effect of the deformation parameter $a$ in the angle $\varphi$ in
which is enclosed the $50$ \% of deflections during the binary encounters,
that is, $\varphi=a\tan\left[  0.5\arctan\left(  \frac{\pi}{a}\right)
\right]  $. Dash line represents the asymptotic value for the uniform
differential cross section (\ref{dCS1}).}%
\label{adcs}%
\end{center}
\end{figure}

The FIG.\ref{adcs} shows the effect of the deformation parameter $a$ on the
angle $\varphi$ in which is enclosed the $50$ \% of the binary encounters,
$\left\vert \theta\right\vert \leq\varphi$. When $a$ is large enough, the
expression (\ref{dCS2}) becomes in a uniform distribution:%

\begin{equation}
d\sigma\left(  \theta\right)  \simeq\sigma\frac{d\theta}{2\pi}, \label{dCS1}%
\end{equation}
which will be used as a reference.%

Let $N_{1}$ be the number of particles remaining in the system. We consider
the $\rho=N_{1}/N$ as the relative density of the gas. We denoted by
$\epsilon$, the kinetic energy per particle of all those particles remaining
in the system, that is:%

\begin{equation}
\epsilon=\frac{1}{N_{1}}\sum_{k=1}^{N}b_{k}\frac{1}{2}\mathbf{v}_{k}^{2} \;
\text{.}%
\end{equation}

The probability that the interaction takes place for a given pair is:%

\begin{equation}
p_{\text{coll}}=\frac{N_{1}\left(  N_{1}-1\right)  }{N\left(  N-1\right)
}\simeq\rho^{2},
\end{equation}
being $N_{1}$ and $N$ large enough. The expectation value of collisions which
take place in a computational step, $c$, can be estimated by $c\simeq
Mp_{\text{coll}}\simeq M\rho^{2}$.

According to the kinetic theory of gases, the rate of collision events per the
time unit, $\omega$, can be estimated as $\omega\simeq\frac{1}{2}\sigma
n^{2}\bar{v}V$, being $\sigma$ the total cross section of the particles, $n$,
the particles per unit of volume, $\bar{v}$, the expectation value of the
relative velocity among the particles of the system, while $V$ is the volume
of the gas container. Let $\Delta t$ be the temporal increment in each
computational step. Since $c\sim\omega\Delta t$, a direct comparison allows us
to set the equivalences: $\rho\sim n$, $N\sim V$, and therefore, the number of
samples is each computational step should grow proportional to the system
size, that is, $M\simeq\mu N$, where $\mu\equiv\frac{1}{2}\sigma n_{0}\bar
{v}\Delta t$, being $n_{0}$ the initial density of particles per unit of
volume. If $m$ represents the particles mass, the characteristics units for
velocity and time are the following: $v_{c}=\sqrt{2u_{c}/m},~\tau_{0}=2/\sigma
n_{0}v_{c},$where $v_{c}$ is also the \textit{cutoff velocity}. The temporal
increment in each computational step $\Delta t$ can be estimated by using the
average relative velocity among the particles of the system as $\Delta
t=\lambda_{0}\mu/\bar{v}$, being:%

\begin{equation}
\bar{v}=\frac{2}{N_{1}\left(  N_{1}-1\right)  }\sum_{k=1}^{N-1}\sum
_{j=k+1}^{N}b_{k}b_{j}\left\vert \mathbf{v}_{k}-\mathbf{v}_{j}\right\vert ,
\end{equation}
and $\lambda_{0}=v_{c}\tau_{0}$, the characteristics particle mean-free-path.

\subsection{Thermodynamical limit}

It could be spoken about a continuous distribution of the particles velocities
when $N$ tends to infinity because of the particles fill densely all the
accessible velocities. Let $n\left(  \mathbf{v},\tau\right)  d^{2}\mathbf{v}$
be the relative density of those particles remaining in the system whose
velocities are located inside the differential region $\left(  \mathbf{v}%
,\mathbf{v}+d\mathbf{v}\right)  $ in a given computational step $\tau$. The
function $n\left(  \mathbf{v},\tau\right)  $ represents the relative density
distribution function of particles in the neighborhood of the point
$\mathbf{v}$ in the space of admissible velocities $\Sigma\equiv\left\{
\mathbf{v}\in\mathbf{R}^{2}:\mathbf{v}^{2}<v_{c}^{2}\right\}  $. The variation
of $n\left(  \mathbf{v},\tau\right)  $ in the next computational step
$\tau+\delta\tau$ can be expressed as follows:%

\begin{equation}
\frac{\delta}{\delta\tau}n\left(  \mathbf{v},\tau\right)  =\mu\left[
I_{in}\left(  \mathbf{v};\tau\right)  -I_{out}\left(  \mathbf{v};\tau\right)
\right]  , \label{conden}%
\end{equation}
being $I_{in}\left(  \mathbf{v};\tau\right)  $ the probability of some
particles to reach a final velocity $\mathbf{v}$ after a collision, while
$I_{out}\left(  \mathbf{v};\tau\right)  $ represents the probability of some
particles with initial velocity $\mathbf{v}$ to experience a collision in the
next computational step. These probabilities are given by the formulae:%

\begin{equation}
I_{in}\left(  \mathbf{v};\tau\right)  =%
{\displaystyle\iint\limits_{\Sigma\times\Sigma}}
d^{2}\mathbf{v}_{1}d^{2}\mathbf{v}_{2}~K\left(  \left.  \mathbf{v}\right\vert
~\mathbf{v}_{1},\mathbf{v}_{2}\right)  n\left(  \mathbf{v}_{1},\tau\right)
n\left(  \mathbf{v}_{2},\tau\right)  ,
\end{equation}%
\begin{equation}
I_{out}\left(  \mathbf{v};\tau\right)  =n\left(  \mathbf{v},\tau\right)
\underset{\Sigma}{\int}d^{2}\mathbf{v}_{1}n\left(  \mathbf{v}_{1},\tau\right)
,
\end{equation}
being $K\left(  \left.  \mathbf{v}\right\vert ~\mathbf{v}_{1},\mathbf{v}%
_{2}\right)  $ the probability that a given particle of the collision pair
with initial velocities $\left(  \mathbf{v}_{1},\mathbf{v}_{2}\right)  $ to
reach $\mathbf{v}$ as a final velocity:%

\begin{equation}
K\left(  \left.  \mathbf{v}\right\vert \mathbf{~v}_{1},\mathbf{v}_{2}\right)
=\int_{0}^{2\pi}~\frac{d\sigma\left(  \theta\right)  }{\sigma}\delta\left[
\mathbf{v-v}\left(  \mathbf{v}_{1},\mathbf{v}_{2};\theta\right)  \right]  ,
\end{equation}
being%

\begin{equation}
\mathbf{v}\left(  \mathbf{v}_{1},\mathbf{v}_{2};\theta\right)  =\frac
{\mathbf{v}_{1}+\mathbf{v}_{2}}{2}+\left\vert \frac{\mathbf{v}_{1}%
-\mathbf{v}_{2}}{2}\right\vert \mathbf{n}\left(  \theta_{0}+\theta\right)  ,
\end{equation}
and $d\sigma\left(  \theta\right)  $ the differential cross section for the
binary encounters. The difference between $I_{in}$ and $I_{out}$ extracts out
all those cases in which the interacting particles keep the same velocities
after the interaction.%

\begin{figure}
[ptb]
\begin{center}
\includegraphics[
height=1.689in,
width=2.3367in
]%
{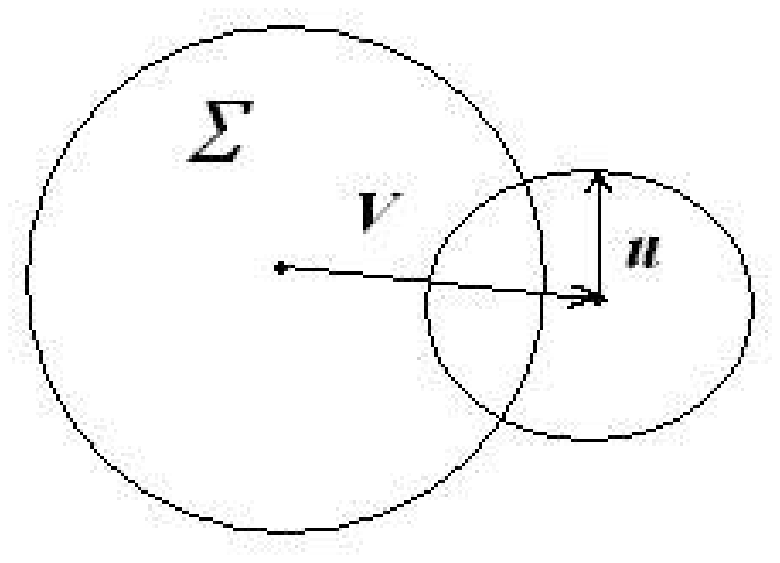}%
\caption{Geometrical visualization of the inequality (\ref{inequal1}). }%
\label{prob}%
\end{center}
\end{figure}

The specific form of the irreversible dynamical equation (\ref{conden}) leads
to some straightforward consequences. Because the probability $p\left(
\mathbf{v}_{1},\mathbf{v}_{2}\right)  $ that a given particle of the collision
pair $\left(  \mathbf{v}_{1},\mathbf{v}_{2}\right)  $ to remain in the system
after the collision:%

\begin{equation}
p\left(  \mathbf{v}_{1},\mathbf{v}_{2}\right)  =\underset{\Sigma}{\int}%
d^{2}\mathbf{v~}K\left(  \left.  \mathbf{v}\right\vert ~\mathbf{v}%
_{1},\mathbf{v}_{2}\right)  ,
\end{equation}
may differ from the unity, $0\leq p\left(  \mathbf{v}_{1},\mathbf{v}%
_{2}\right)  \leq1,$ the total relative density of the system $\rho\left(
\tau\right)  =\int_{\Sigma}d^{2}\mathbf{v~}n\left(  \mathbf{v},\tau\right)  $
will decrease or remain constant:%

\begin{equation}
\frac{\delta}{\delta\tau}\rho\left(  \tau\right)  =-\mu%
{\displaystyle\iint\limits_{\Sigma\times\Sigma}}
d^{2}\mathbf{v}_{1}d^{2}\mathbf{v}_{2}~q\left(  \mathbf{v}_{1},\mathbf{v}%
_{2}\right)  n\left(  \mathbf{v}_{1},\tau\right)  n\left(  \mathbf{v}_{2}%
,\tau\right)  \leq0, \label{inequal1}%
\end{equation}
being $q\left(  \mathbf{v}_{1},\mathbf{v}_{2}\right)  =1-p\left(
\mathbf{v}_{1},\mathbf{v}_{2}\right)  $, the escaping probability.

A geometrical visualization of the inequality (\ref{inequal1}) is shown in
FIG.\ref{prob}. Here $\mathbf{\Sigma}$ is the admissible subset of velocities
in $\mathbf{R}^{2}$; $\mathbf{V}$ represents the velocity of the center of
mass, $\mathbf{V}=\frac{1}{2}\left(  \mathbf{v}_{1}+\mathbf{v}_{2}\right)  $,
while $\mathbf{u}$ is the final velocity of the specific particle in the
center of mass frame, that is, $\mathbf{u}=\frac{1}{2}\left\vert
\mathbf{v}_{1}-\mathbf{v}_{2}\right\vert \mathbf{n}\left(  \theta_{0}%
+\theta\right)  $. Only those values of the vector addition $\mathbf{v}%
=\mathbf{V+u}$ belonging to $\mathbf{\Sigma}$ correspond to the final state
where the specific particle of the pair remains in the system after the
collision. In the particular case of the uniform differential cross section
(\ref{dCS1}), $p\left(  \mathbf{v}_{1},\mathbf{v}_{2}\right)  $ is numerically
equal to fraction of the circumference perimeter described by the vector
$\mathbf{u}$ belonging to the subset $\mathbf{\Sigma}$. A simple geometrical
analysis yields the following expression for this probability:%

\begin{equation}
p\left(  \mathbf{v}_{1},\mathbf{v}_{2}\right)  =\left\{
\begin{array}
[c]{cc}%
0 & \text{when }x\leq-1\\
\frac{1}{2}+\frac{1}{\pi}\arcsin\left(  x\right)  & \text{when }\left\vert
x\right\vert <1\\
1 & \text{when }x\geq1
\end{array}
\right\}  ,
\end{equation}
where $x$ is given by:%

\begin{equation}
x=\frac{v_{c}^{2}-\mathbf{V}^{2}-\mathbf{u}^{2}}{2\left\vert \mathbf{V}%
\right\vert \left\vert \mathbf{u}\right\vert }\;\text{.}%
\end{equation}

It is evident that only those collision pairs satisfying the condition:%

\begin{equation}
\frac{1}{2}\left\vert \mathbf{v}_{1}+\mathbf{v}_{2}\right\vert +\frac{1}%
{2}\left\vert \mathbf{v}_{1}-\mathbf{v}_{2}\right\vert \leq v_{c},
\label{condition}%
\end{equation}
will remain in the system after the interaction with a $100$ \% of
probability, and this condition does not depend on the differential cross
section used to describe the binary collisions, nor on the dimension of the space.

As already shown, the relative density distribution function $n\left(
\mathbf{v};\tau\right)  $ obeys a nonlinear irreversible dynamics, which is
also \textit{dissipative}. Thus, the evaporation is \textit{explicitly taken
into account} in the dynamical equation of the system. The nonlinear nature of
equation (\ref{conden}) is a feature making the difference from the standard
treatment of evaporation by using a linear Fokker-Planck equation applied to
the astrophysical systems \cite{bin,chandra,spitzer,king}. This equation does
not necessarily lead to exponential decays of the particles number with the time.

\subsection{Numerical experiments}

We set without lost of generality $m=1$, $\lambda_{0}=1$ and $u_{c}=2.6$ in
our numerical study of dynamics. The number $M$ of interacting pair chosen in
a given computational step was settled to be $M=\left[  \mu N\right]  $, where
$\mu=0.05$, in order to ensure a smooth temporal evolution of the physical
observables. We start from an initial configuration where all particles belong
to the system, that is, $b_{k}=1$, $\forall k\in\left[  1,2,...N\right]  $.
Preliminary runs by using the uniform differential cross section (\ref{dCS1})
allows us to set $N=1000$, since size effects are not significant when we
compare the results obtained by considering $N=500$ and $1000$.

We are interested in describing some kind of dynamical evolution of the model
system which could be considered as a quasi-stationary regime. To this aim we
take into account the following initial distribution functions of particle energies:

\begin{itemize}
\item Uniform Distribution (\textbf{UNIC}):%
\begin{equation}
\omega_{I}\left(  e\right)  =\frac{1}{E_{1}}\times\left\{
\begin{array}
[c]{cc}%
1 & \text{with }0\leq e<E_{1}<u_{c}\\
0 & \text{otherwise}%
\end{array}
\right.  , \label{UNIC}%
\end{equation}

\item Truncating Boltzmann Distribution (\textbf{TBIC}):%
\begin{equation}
\omega_{II}\left(  e\right)  =\frac{\beta\exp\left(  -\beta e\right)  }%
{1-\exp\left(  -\beta u_{c}\right)  }\times\left\{
\begin{array}
[c]{cc}%
1 & \text{ with }0\leq e<u_{c}\\
0 & \text{otherwise}%
\end{array}
\right.  , \label{TBIC}%
\end{equation}

\item Double Size Uniform Distribution (\textbf{DSIC}):
\end{itemize}

\begin{equation}
\omega_{III}\left(  e\right)  =\frac{1}{E_{2}-E_{1}}\times\left\{
\begin{array}
[c]{cc}%
1 & \text{with }E_{1}<e<E_{2}<u_{c}\\
0 & \text{otherwise}%
\end{array}
\right.  , \label{DSIC}%
\end{equation}
where $E_{1}$, $E_{2}$ and $\beta$ are positive real parameters. The particles
velocities were uniformly distributed at random in all possible directions
$\left(  0,2\pi\right)  $ for the three initial conditions.

\begin{figure}
[t]
\begin{center}
\includegraphics[
height=3.039in,
width=3.2396in
]%
{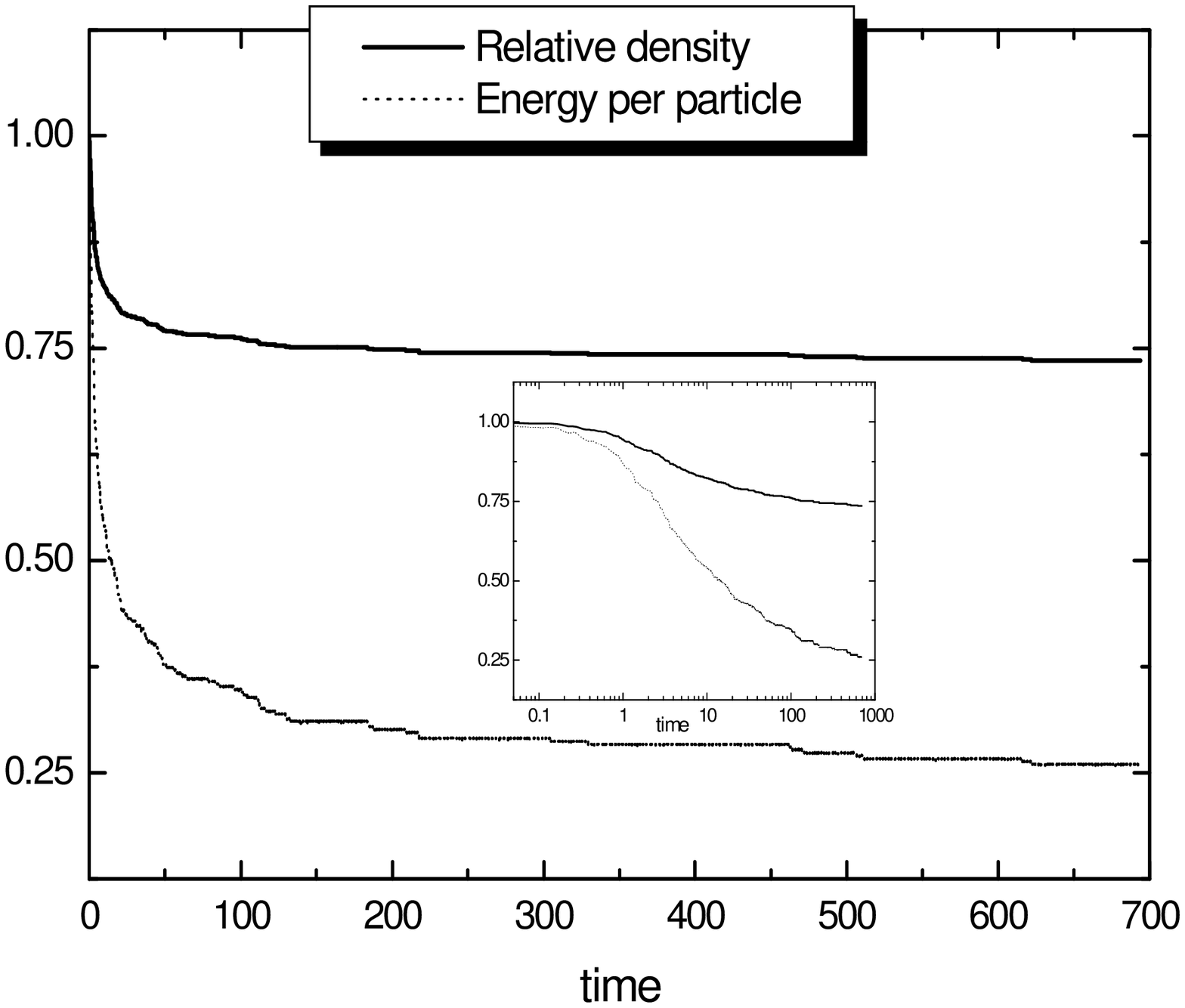}%
\caption{Dynamical evolution of the relative density $\rho$ and the kinetic
energy per particle $\epsilon$ by using the uniform differential cross section
with \textbf{UNIC} initial condition with $E_{1}=2.0$. The same dependences
are shown in the insert graph by using a logarithmic scale for the time
variable.}%
\label{dcs1a}%
\end{center}
\end{figure}

We start from an initial condition \textbf{UNIC} with $E_{1}=2.0$ by using the
uniform differential cross section. FIG.\ref{dcs1a} shows the dynamical
evolution of $\rho$ and $\epsilon$ in terms of the physical time $t$. It is
easily seen that the system experiences a sudden relaxation process in which
the $25$ \% of the particles is evaporated in the first stages of the
evolution. There, the system enters in a long transient regime where the
evaporation events have been extremely reduced, but they still take place
causing a slow variation of the kinetic energy per particle.

The insert graph evidences that this relaxation process is progressive, that
is, no real equilibrium configuration seems to be reached by extending the
computational simulation of the dynamics, but the rate of the evolution seems
to be continuously reduced.

Let us analyze now the effect of the particle evaporation in the energy
distribution function of the system. We will apply the following procedure in
order to obtain an appreciable statistics of the microscopic configurations
around a given computational step $\tau$: \ We carry out certain number $v$ of
virtual trajectories, with $100$ computational steps each, by starting from
the initial configuration at the computational step $\tau$. The data is
collected by considering $20$ microscopic configurations in each of such
virtual trajectories, and therefore, a total of $20v$ microscopic
configurations are taken into account in order to build the energy
distribution function. This procedure is applied in the successive analysis.

\begin{figure}
[t]
\begin{center}
\includegraphics[
height=3.039in,
width=3.2396in
]%
{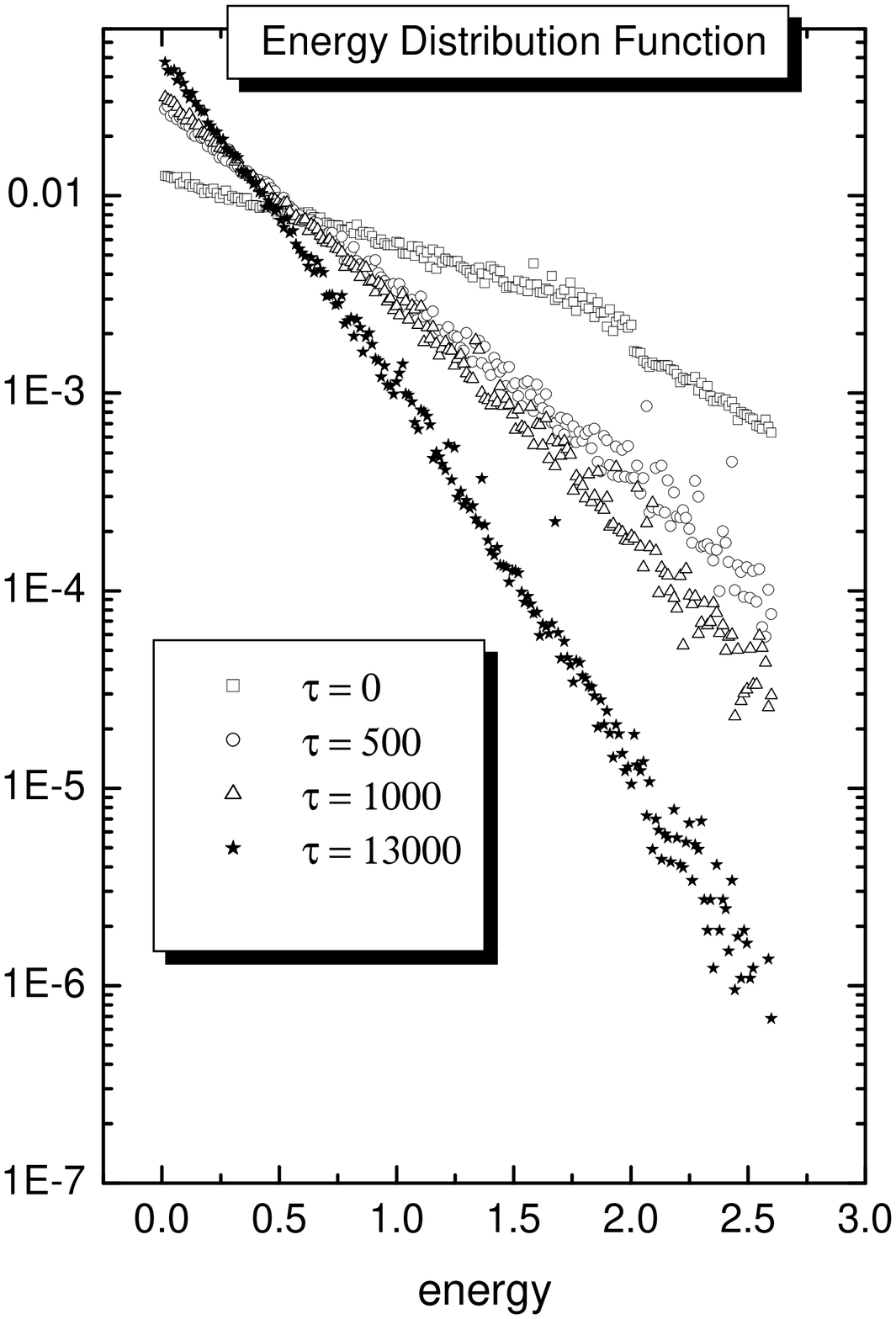}%
\caption{Energy distribution function along the dynamical evolution shown in
FIG.\ref{dcs1a}. The system seems to stabilize in a\ Truncating\ Boltzmann
distribution.}%
\label{dcs1adf}%
\end{center}
\end{figure}

FIG.\ref{dcs1adf} shows the energy distribution function by considering $100$
virtual trajectories: at the first stages of the evolution with $\tau
=0,500,1000$, and a final configuration when $\tau=13000$. It is remarkable
that after a very short relaxation, the energy distribution function is
stabilized in a truncating Boltzmann distribution of the form given in the
equation (\ref{TBIC}), which is easily evidenced in the linear dependence of
the distribution functions shown in this figure by using a logarithmic scale.
The existence of this profile allows us to claim that the system exhibits a
\textit{quasi-stationary evolution} where the relative density $\rho$ and the
energy per particle $\epsilon$ seem to be the macroscopic control variables
along the successive dynamics of the system.%

%

\begin{figure}
[t]
\begin{center}
\includegraphics[
height=3.2396in,
width=3.2396in
]%
{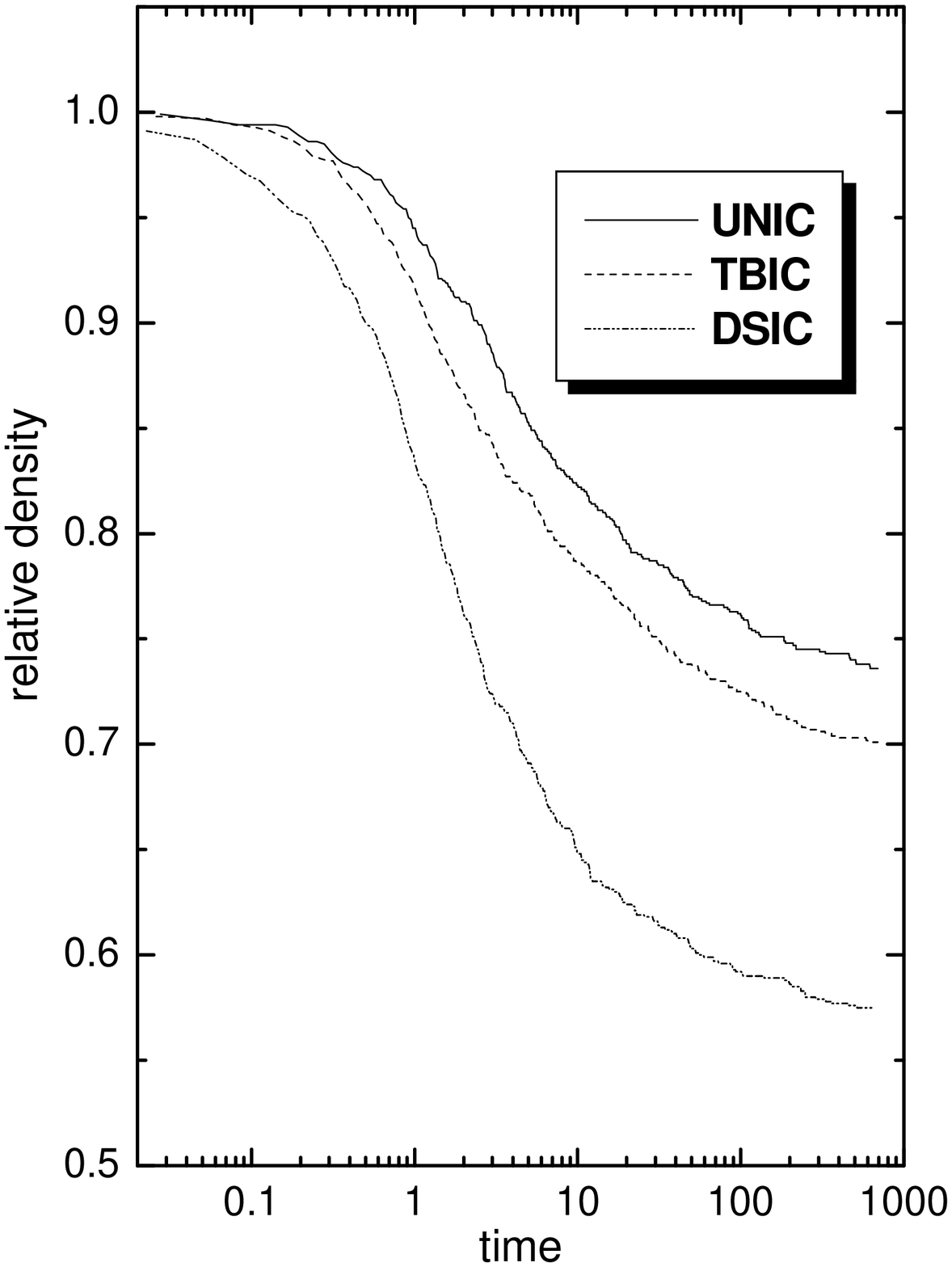}%
\caption{Dynamical evolution of the relative density by using three different
initial conditions and the uniform differential cross section.}%
\label{dcs1comd}%
\end{center}
\end{figure}

\begin{figure}
[t]
\begin{center}
\includegraphics[
height=3.1903in,
width=3.2396in
]%
{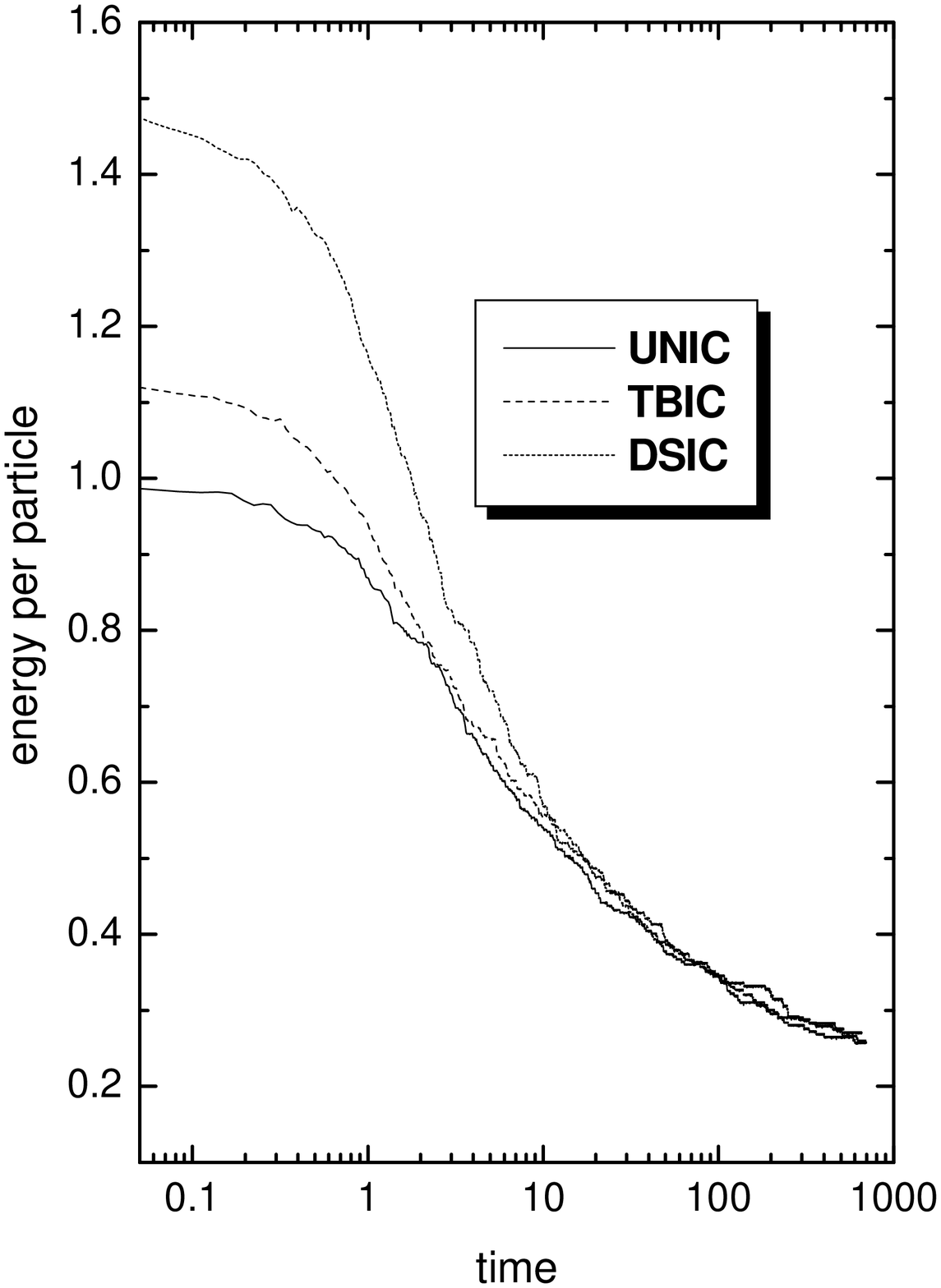}%
\caption{Dynamical evolution of the energy per particle by using three
different initial conditions and the uniform differential cross section.}%
\label{dcs1com}%
\end{center}
\end{figure}

Let us now perform a comparative study in order to clarify the possible
influence of the initial conditions in the nature of the quasi-stationary
evolution of the system. We consider three initial conditions for this
analysis: $a$) \textbf{UNIC}: $E_{1}=2.0$; $b$) \textbf{TBIC}: $\beta=0.25$;
$c$) \textbf{DSIC}: $E_{1}=1.0$ and $E_{2}=2.0$. The results are shown in
FIG.\ref{dcs1comd} and FIG.\ref{dcs1com}. As expected, the particle
evaporation depends strongly on the initial microscopic configurations.
However, the energy per particle $\epsilon$ quickly tends to a regime where
the successive evolution is so slow that the dynamical evolutions of this
observable starting from different initial microscopic configurations converge
asymptotically all them.

\begin{figure}
[t]
\begin{center}
\includegraphics[
height=3.2396in,
width=3.2396in
]%
{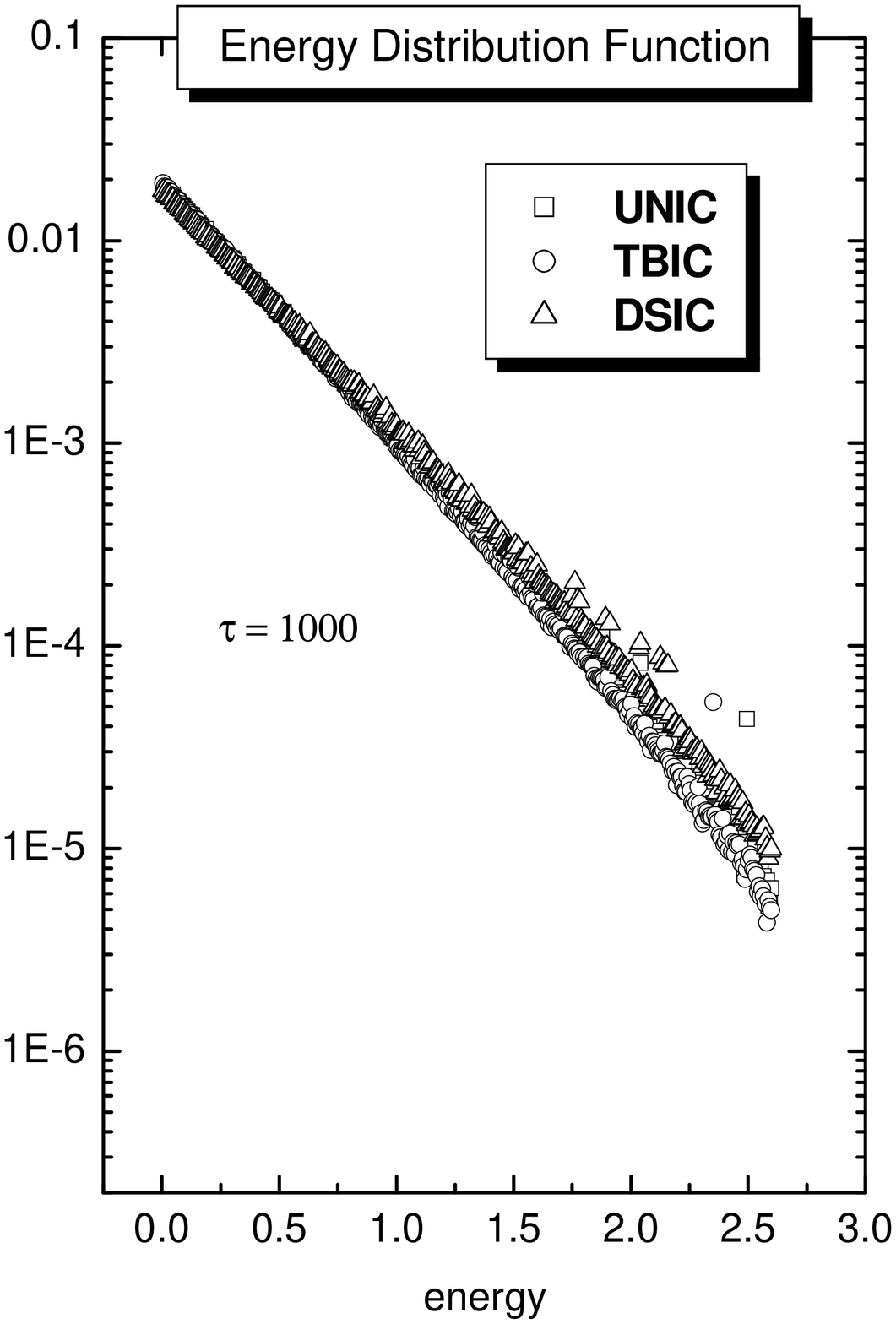}%
\caption{Energy distribution function at $\tau=1000$ for the dynamical
evolution shown in FIG.\ref{dcs1comd} and FIG.\ref{dcs1com}. }%
\label{dcs1comdfc}%
\end{center}
\end{figure}

\begin{figure}
[t]
\begin{center}
\includegraphics[
height=3.2396in,
width=3.2396in
]%
{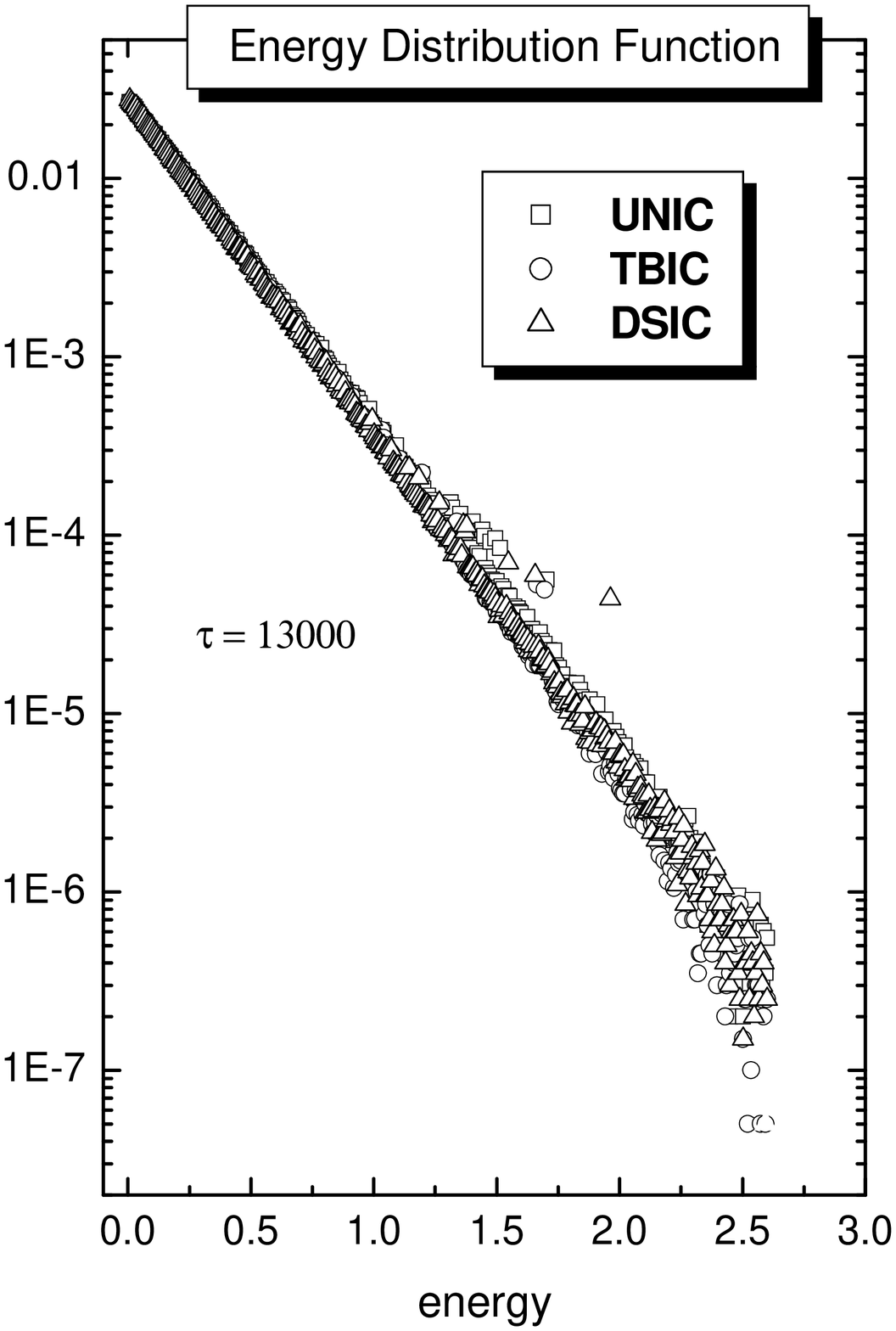}%
\caption{The same observable shown in FIG.\ref{dcs1comdfc}, but this time at
$\tau=13000$. }%
\label{dcs1comdfz}%
\end{center}
\end{figure}

FIG.\ref{dcs1comdfc} and FIG.\ref{dcs1comdfz} show the energy distribution
functions for these initial conditions at two different computational times by
considering also $100$ virtual trajectories: FIG.\ref{dcs1comdfc} at
$\tau=1000$, and FIG.\ref{dcs1comdfz} at $\tau=13000$. These results strongly
suggest us that the Truncating Boltzmann distribution (\ref{TBIC}) is the
\textit{quasi-stationary profile} appearing in the dynamical evolution of a
gas of particles driven by binary encounters which undergoes particle
evaporation by using the uniform differential cross section. The great
similarity among these profiles is a consequence of the sudden evolution of
the system towards configurations whose energy per particle is around
$\epsilon\simeq0.27$. For this energy the evaporation rate is so low that the
subsequent dynamics spends too much time for exhibiting a significant variation.

All those interacting particles whose kinetic energies are less than $u_{c}/2$
do not provoke evaporation events, and therefore evaporation events could be
only produced when a particle, whose energy $e$ belongs to the energetic range
$\frac{1}{2}u_{c}\leq e\leq u_{c}$, is involved in a binary encounter. This is
precisely the origin of the slow dynamical evolution observed in the above
numerical experiment when $\epsilon\simeq0.27$: Most of the particles
remaining in the system belong to the stable region $0\leq e\leq\frac{1}%
{2}u_{c}$.%

%

%

%

%

\begin{figure}
[t]
\begin{center}
\includegraphics[
height=3.1903in,
width=3.2396in
]%
{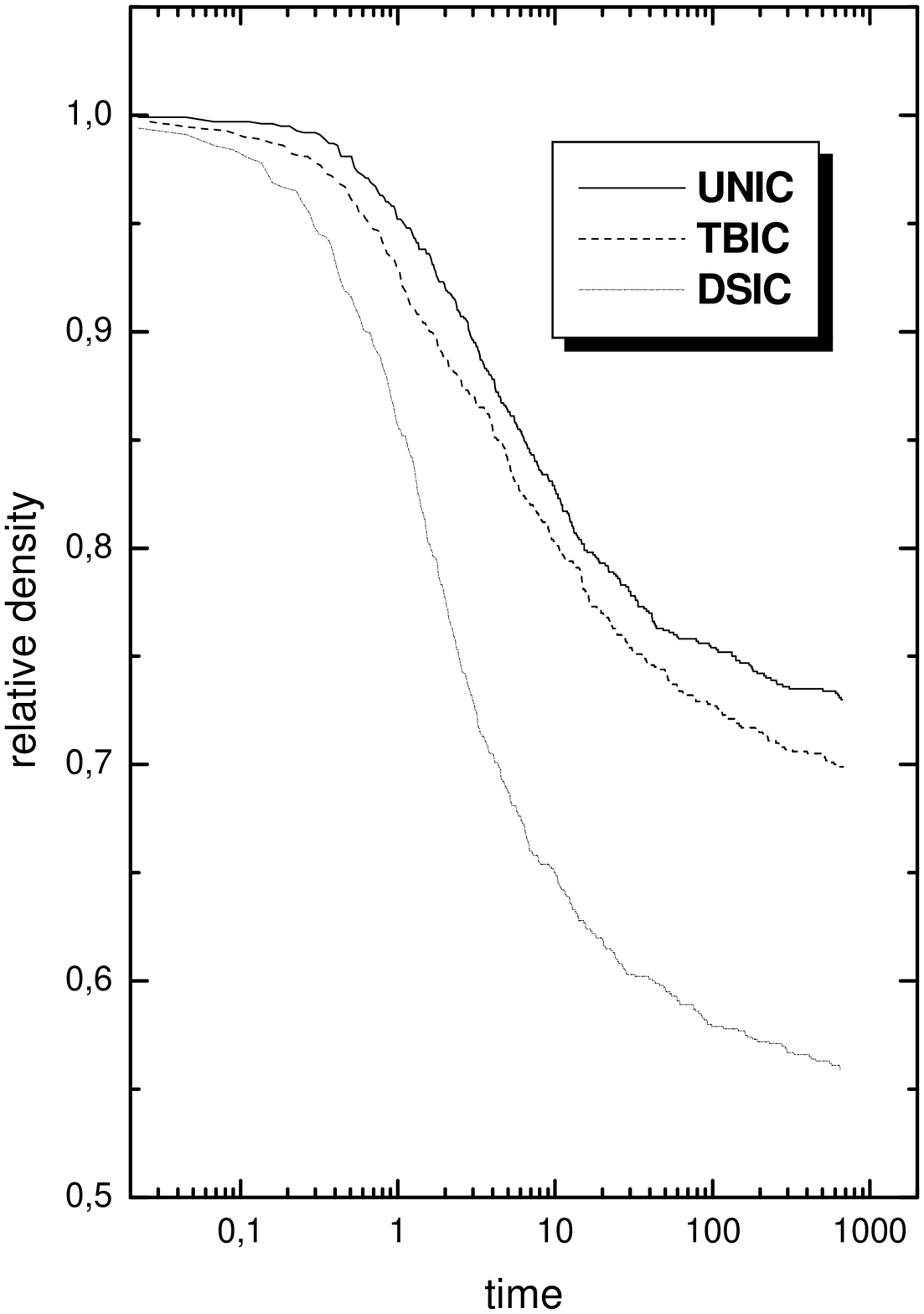}%
\caption{Dynamical evolution of the relative density $\rho$ when a moderate
asymmetry among large and short deflections is considered.}%
\label{dcs2comd}%
\end{center}
\end{figure}
%

\begin{figure}
[t]
\begin{center}
\includegraphics[
height=3.2396in,
width=3.2396in
]%
{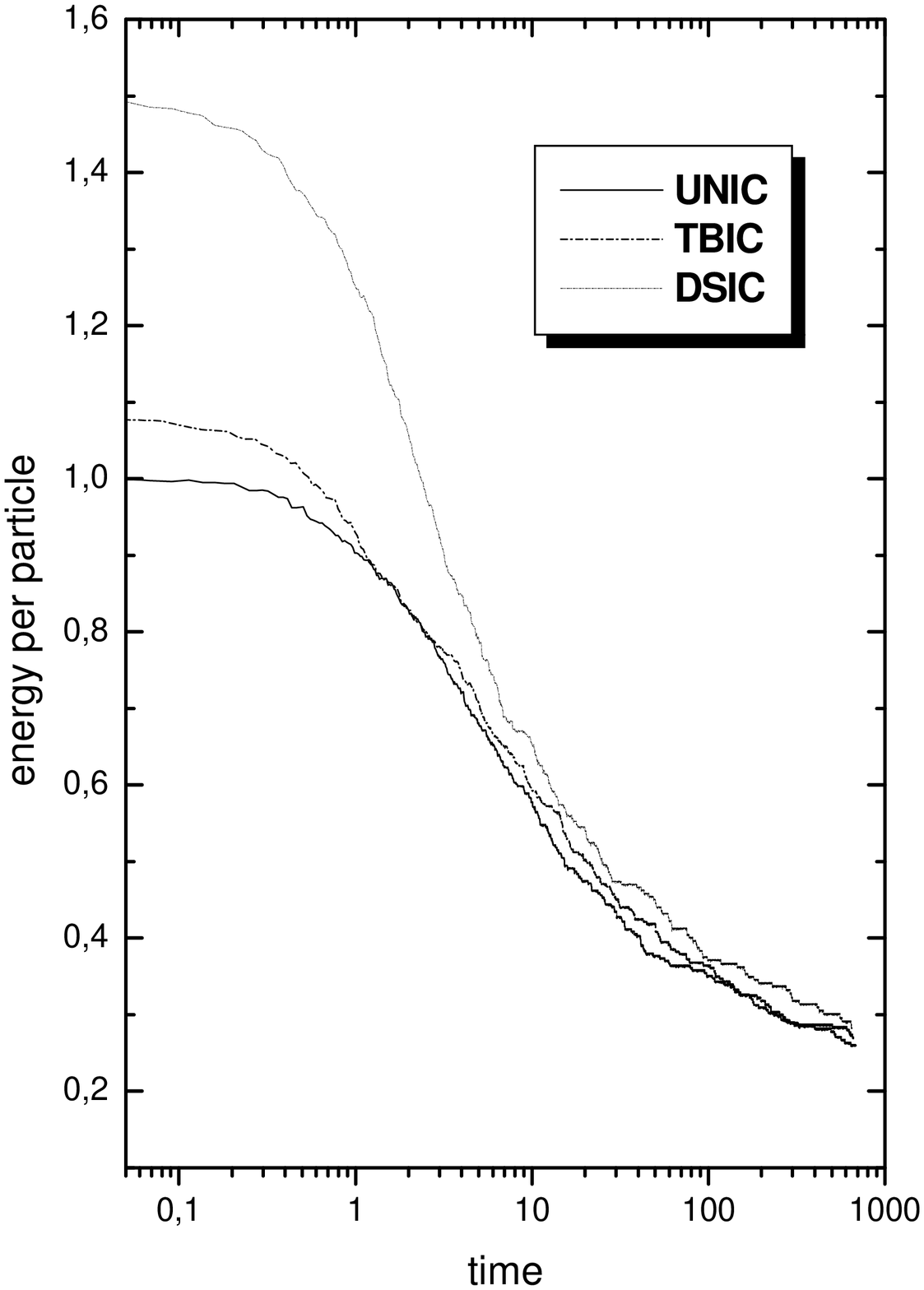}%
\caption{Dynamical evolution of the energy per particle $\epsilon$ by using a
moderate asymmetry among large and short deflections.}%
\label{dcs2com}%
\end{center}
\end{figure}
%

\begin{figure}
[t]
\begin{center}
\includegraphics[
height=3.2396in,
width=3.2396in
]%
{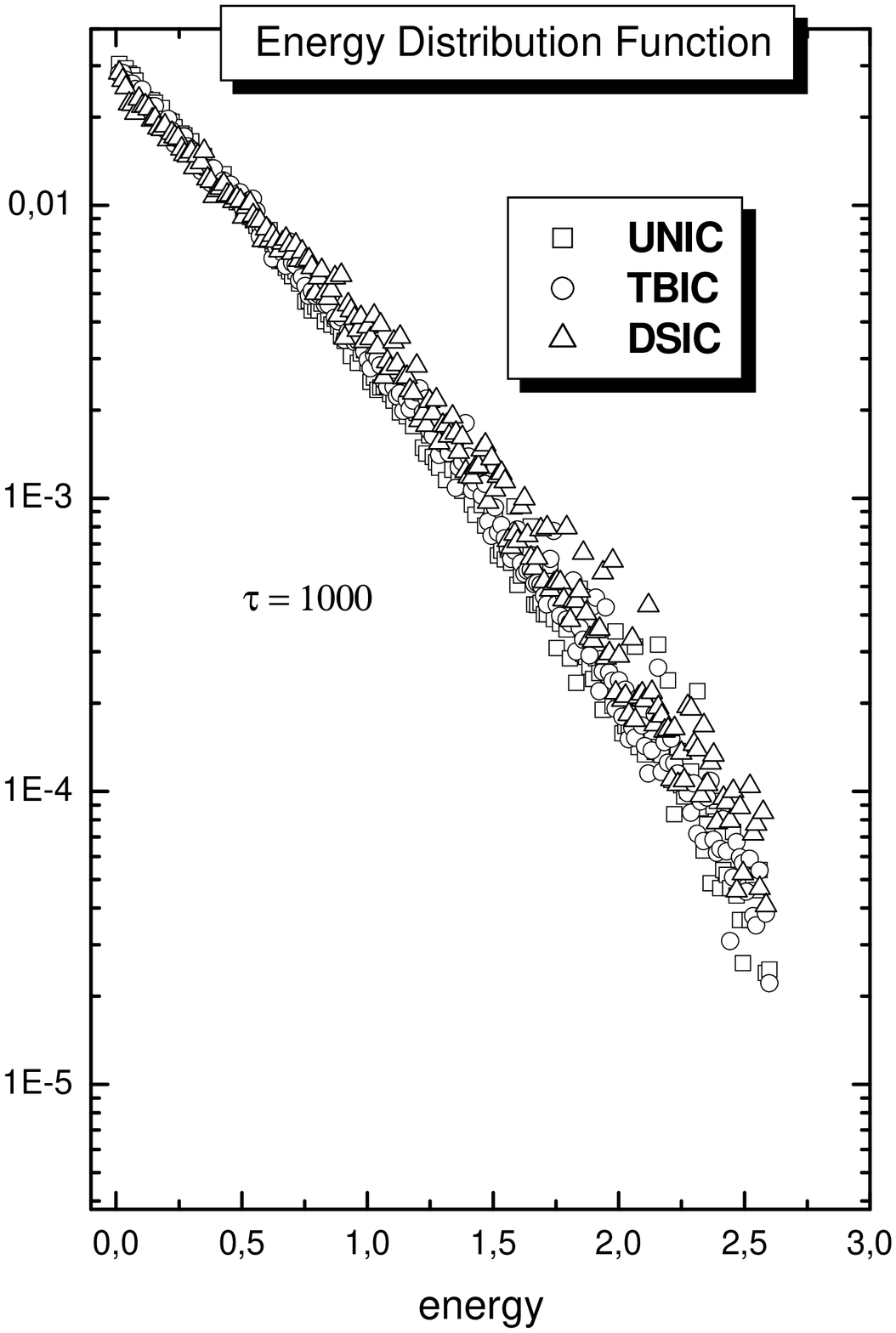}%
\caption{Energy distribution function at $\tau=1000$ for the dynamical
evolutions shown in FIG.\ref{dcs2comd} and FIG.\ref{dcs2com}. }%
\label{dcs2comdfc}%
\end{center}
\end{figure}
%

\begin{figure}
[t]
\begin{center}
\includegraphics[
height=3.2396in,
width=3.2396in
]%
{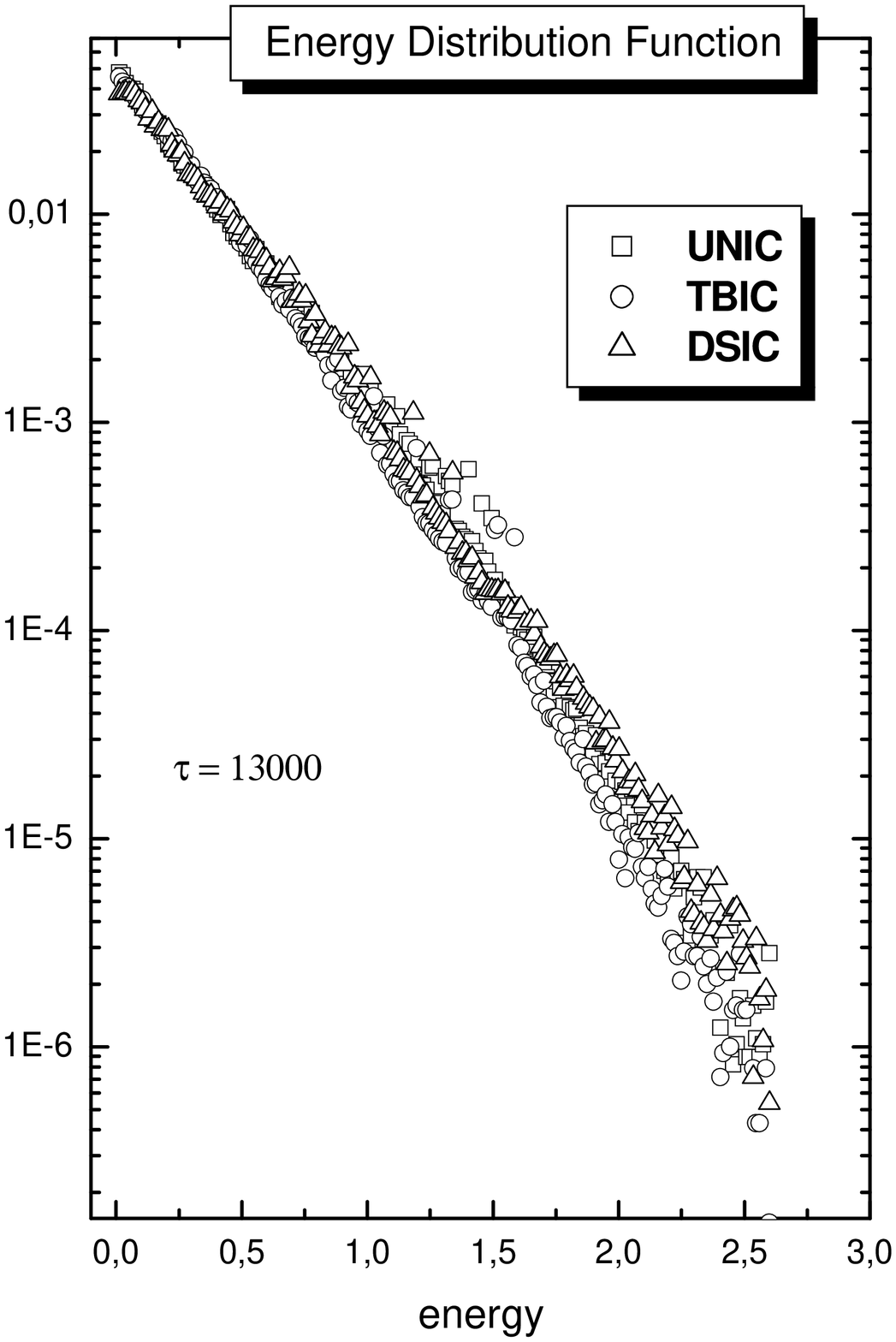}%
\caption{The same observable shown in FIG.\ref{dcs2comdfc}, but this time at
$\tau=13000$. }%
\label{dcs2comdfz}%
\end{center}
\end{figure}

As already noted, the uniform differential cross section (\ref{dCS1}) allows a
given interacting pair to explore; in an effective fashion, all those possible
results of their collision event, and therefore, large variations of the
kinetic energies of the interacting particles are very likely. We are
interested in analyzing if the quasi-stationary profile obtained above remains
unchanged even when the collision events are not so effective, that is, when
large variations of the kinetic energy of the collision particles are not so
favored. This analysis can be performed by using the differential cross
section (\ref{dCS2}) with $a=0.2\pi$. As already shown in FIG.\ref{adcs}, this
particular value produces a moderate asymmetry among large and short
deflections angles during the binary encounters, and therefore, a weaker
effectiveness of the equilibration mechanism with regard to the uniform case.%

\begin{figure}
[t]
\begin{center}
\includegraphics[
height=3.1903in,
width=3.2396in
]%
{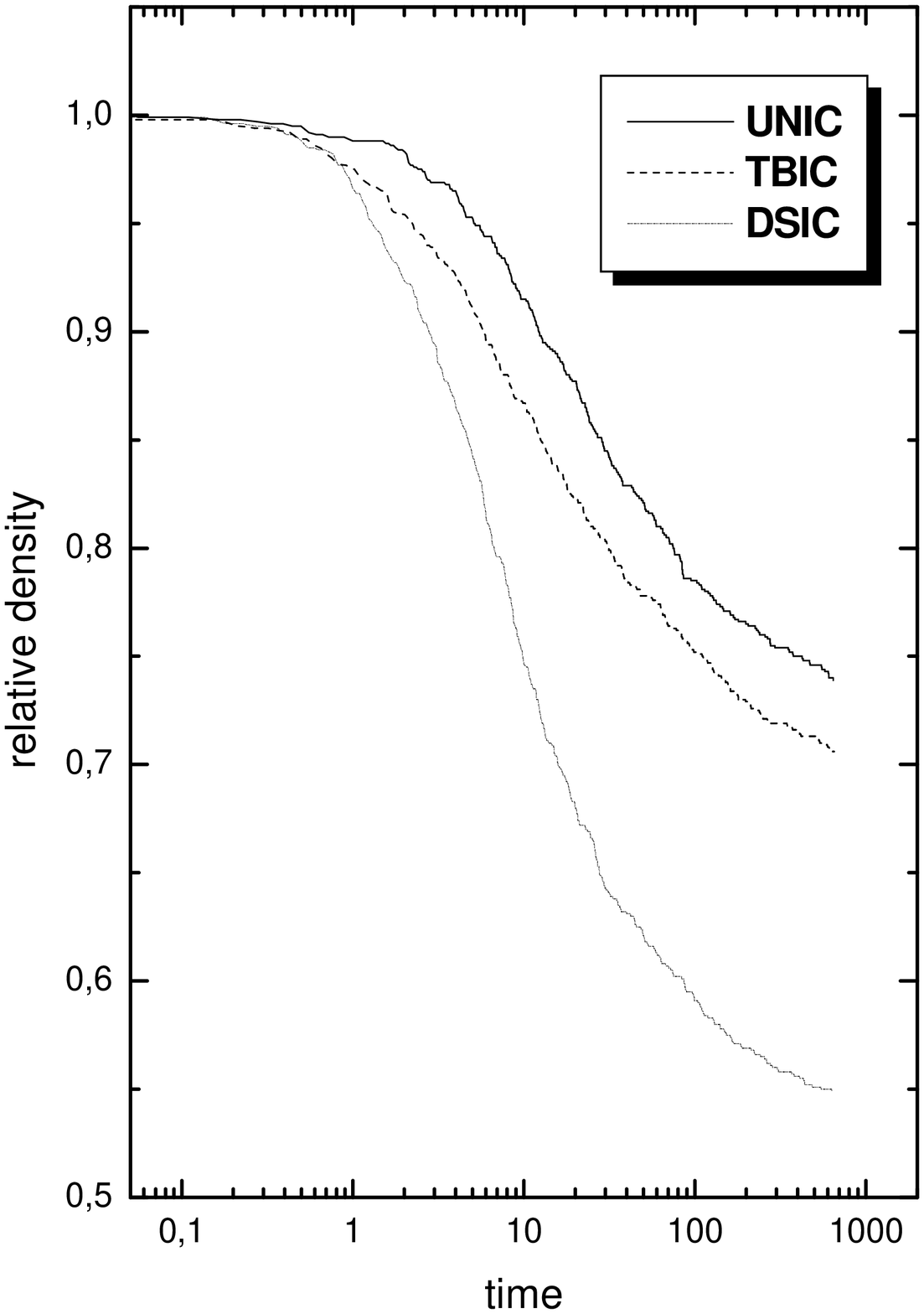}%
\caption{Dynamical evolution of the relative density $\rho$ by using a
differential cross section with a great asymmetry\ between short and large
deflections. }%
\label{dcs3comd}%
\end{center}
\end{figure}
%

\begin{figure}
[t]
\begin{center}
\includegraphics[
height=3.1903in,
width=3.2396in
]%
{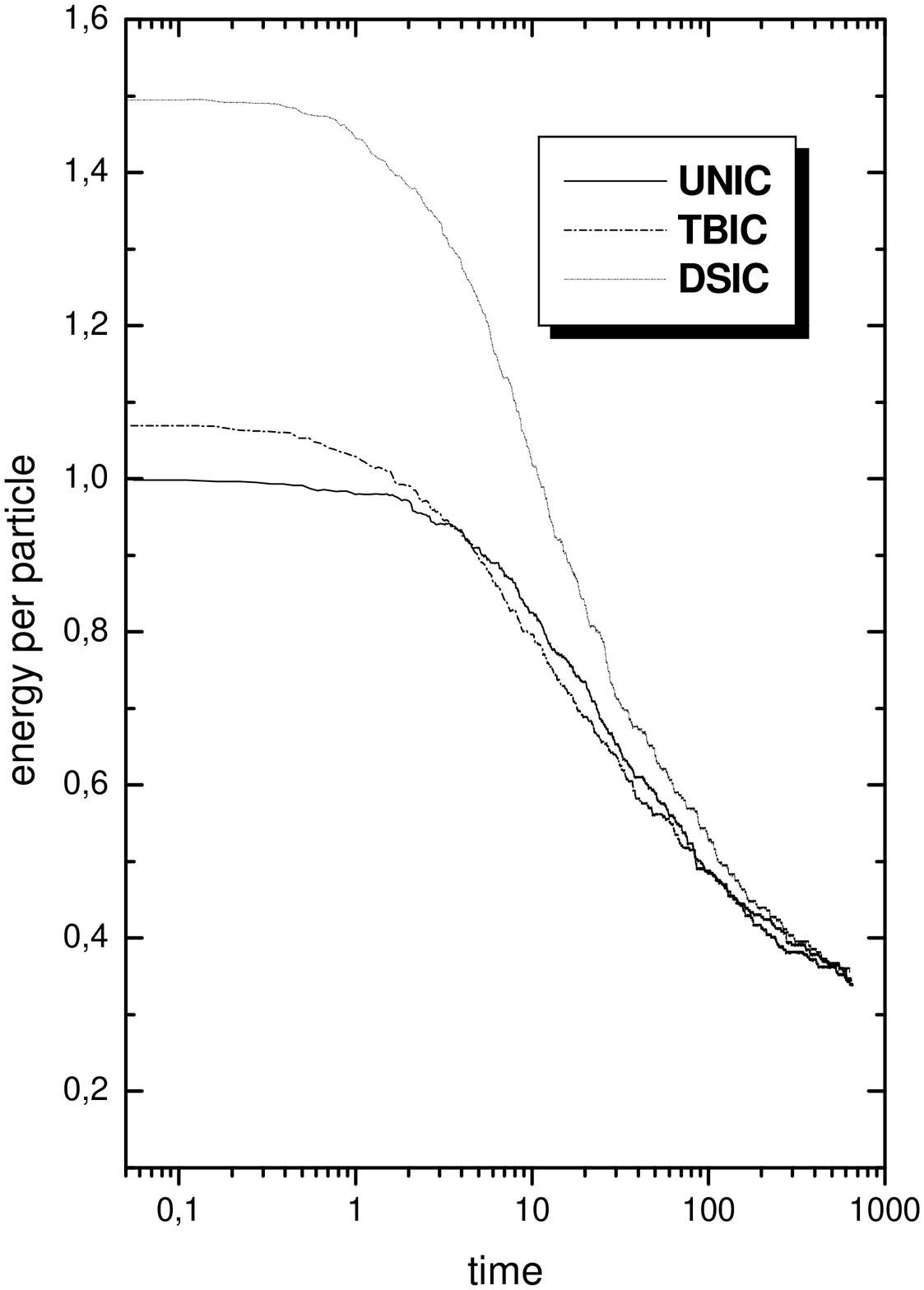}%
\caption{Dynamical evolution of the energy per particle $\epsilon$ by using a
differential cross section with a great asymmetry\ between short and large
deflections. }%
\label{dcs3com}%
\end{center}
\end{figure}
%

\begin{figure}
[t]
\begin{center}
\includegraphics[
height=3.2396in,
width=3.2396in
]%
{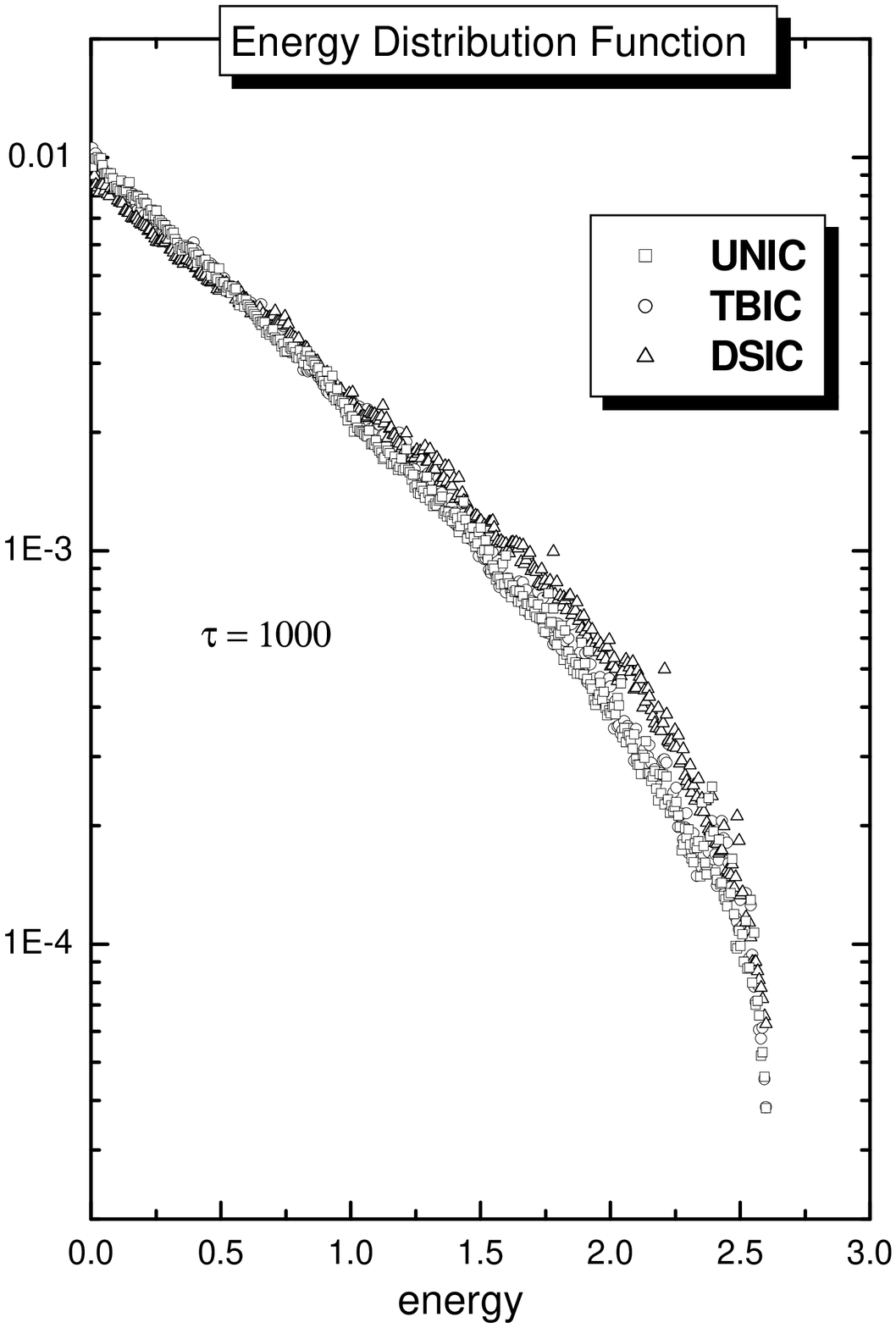}%
\caption{Energy distribution function at $\tau=1000$ for the dynamical
evolutions shown in FIG.\ref{dcs3comd} and FIG.\ref{dcs3com}. }%
\label{dcs3comdfc}%
\end{center}
\end{figure}
%

\begin{figure}
[t]
\begin{center}
\includegraphics[
height=3.2396in,
width=3.2396in
]%
{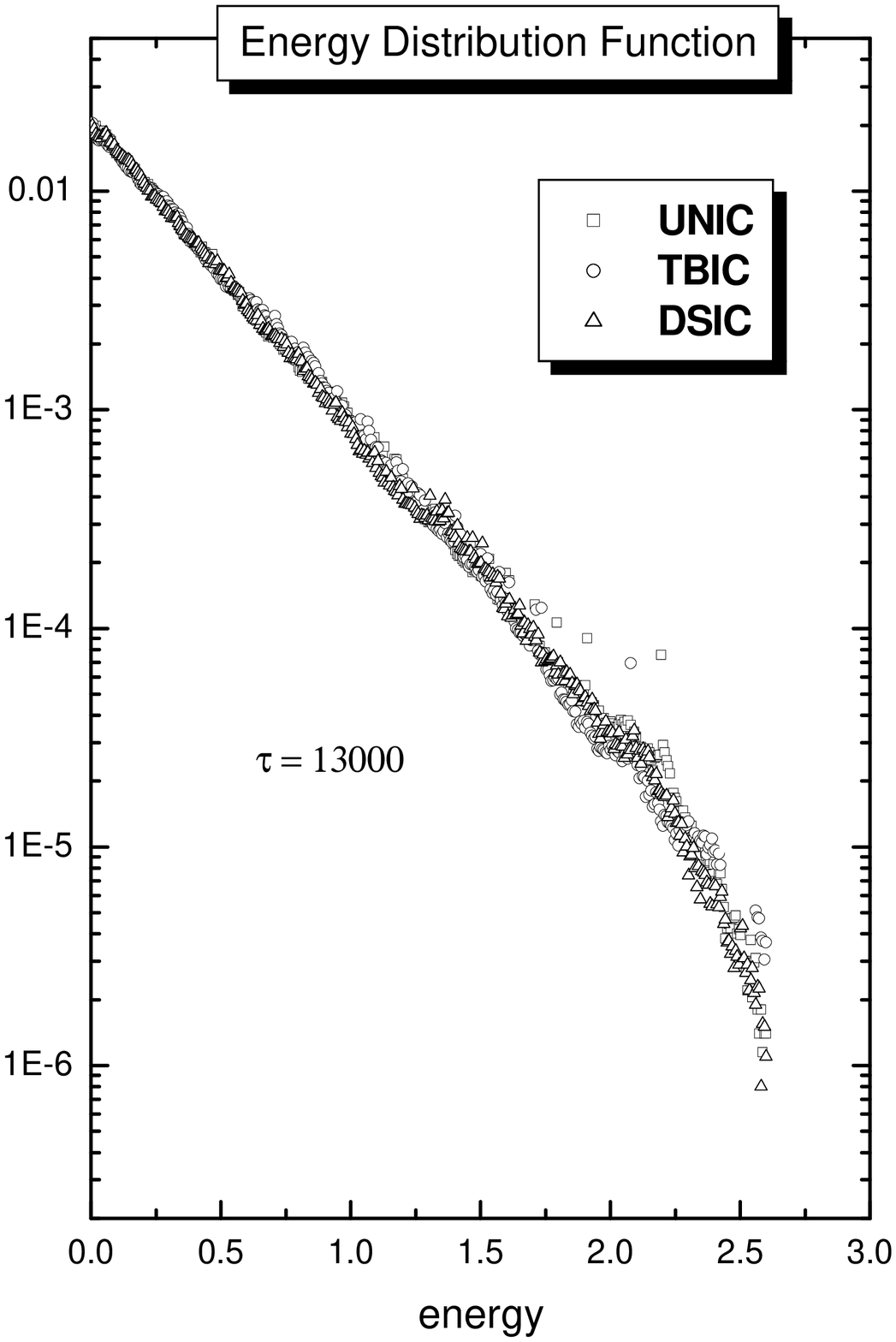}%
\caption{The same observable shown in FIG.\ref{dcs3comdfc}, but this time at
$\tau=13000$. }%
\label{dcs3comdfz}%
\end{center}
\end{figure}

We implement the same initial conditions used in the previous study. The
dynamical evolution of the relative density $\rho$ and the energy per particle
$\epsilon$ are shown in FIG.\ref{dcs2comd} and FIG.\ref{dcs2com},
respectively. Despite we have considered a new microscopic interacting picture
for the binary encounters, it is evident here that no significant qualitative
differences are exhibited in the dynamical behavior of these observables. This
conclusion is reinforced by looking at the results showed in
FIG.\ref{dcs2comdfc} and FIG.\ref{dcs2comdfz}, where it can be appreciated the
persistence of the truncating Boltzmann profile (\ref{TBIC}) even after using
a moderate asymmetry among large and short deflections, although there is a
slight tendency to a reduction of the particles population for the large
energies with a significant incidence of fluctuations.

Let us now consider a very asymmetric differential cross section. We repeat
the same numerical experiment described above, but this time by considering
$a=0.02\pi$. This value of the deformation parameter $a$ leads to a
predominance of the small variation of the kinetic energies of the collision
particles, and therefore, a significant reduction of the effectiveness of the
microscopic dynamics in reaching their quasi-equilibrium conditions.

The evolution of $\rho$ and $\epsilon$ are shown in FIG.\ref{dcs3comd} and
FIG.\ref{dcs3com}. In a similar way, the energy distribution functions when
$\tau=1000$ and $13000$, in FIG.\ref{dcs3comdfc} and FIG.\ref{dcs3comdfz} is
also shown after running 200 virtual trajectories. As expected, the rate of
evaporation has been reduced. The energy distribution function clearly
evidences a new qualitative effect of the evaporation, which is more
accentuated in FIG.\ref{dcs3comdfc}. The results shown in these figures
evidence a significant deviation of the distribution function from the linear
dependence close of the energy cutoff $u_{c}$, \textit{without} vanishing at
this energy.

Let us now investigate, qualitatively, these quasi-stationary profiles by
using a Michie-King-like distribution \cite{bin}:%

\begin{equation}
\omega_{M-K}\left(  e\right)  =C_{0}\langle\exp\left[  -\beta e\right]
-\exp\left(  -\beta\varepsilon_{C}\right)  \rangle, \label{michie}%
\end{equation}
with $0\leq e\leq u_{c}$, being $u_{c}\leq\varepsilon_{C}<+\infty$. It is easy
to note that when $\varepsilon_{C}$ decreases from the infinity towards the
cutoff energy $u_{c}$, the energy parameter $\varepsilon_{C}$ causes a
continuous deformation of the Truncating Boltzmann profile (\ref{TBIC})
towards the well-known Michie-King profile of the globular clusters
\cite{bin,king}.%

\begin{figure}
[t]
\begin{center}
\includegraphics[
height=3.039in,
width=3.2448in
]%
{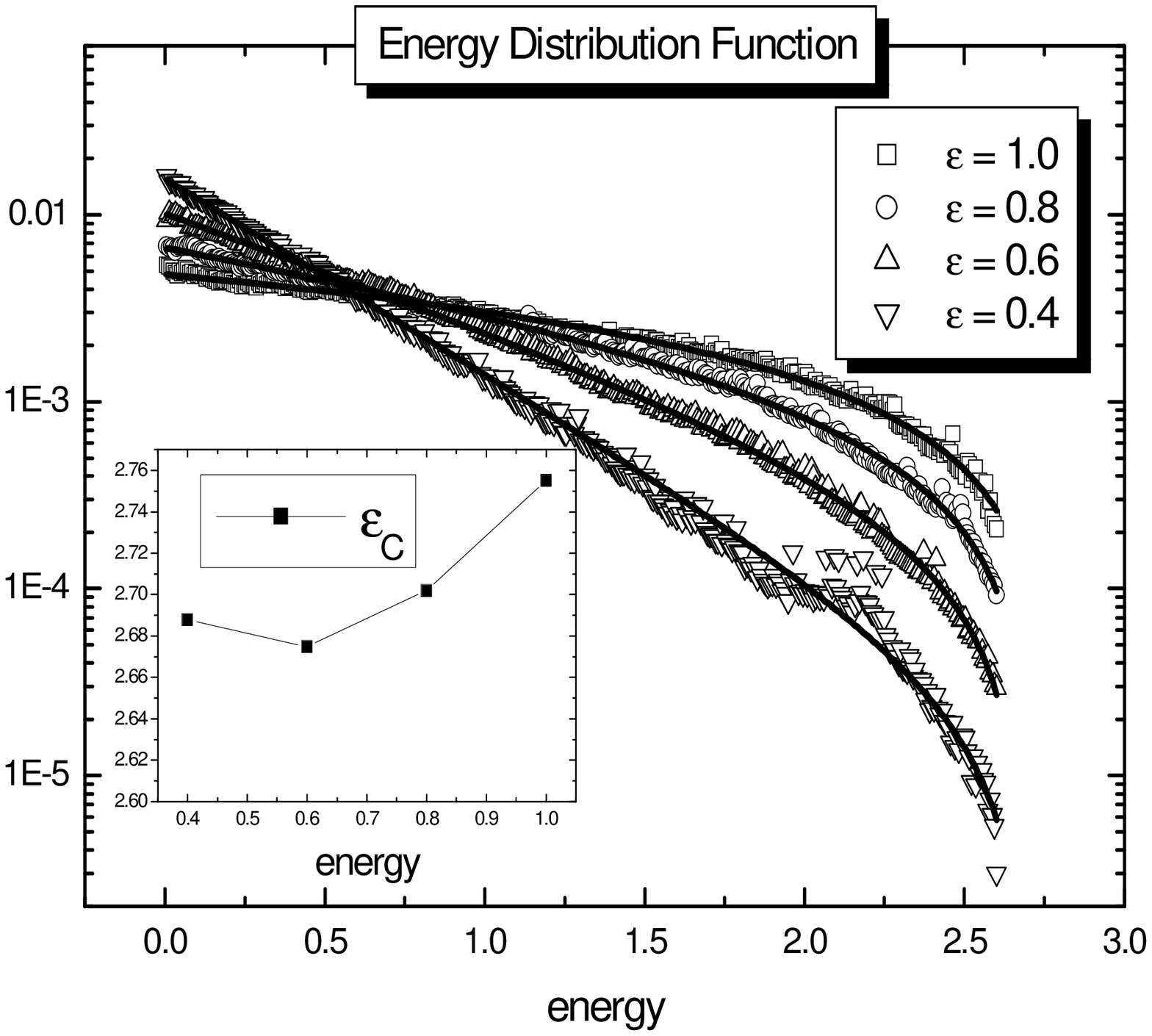}%
\caption{Energy Distribution function with $a=0.02\pi$ at different energy per
particles with their fitting by using the Michie-King like profile. The insert
plot shows the energy dependence of the parameter $\varepsilon_{C}$. }%
\label{dcs3michie}%
\end{center}
\end{figure}
%

\begin{figure}
[t]
\begin{center}
\includegraphics[
height=3.039in,
width=3.2396in
]%
{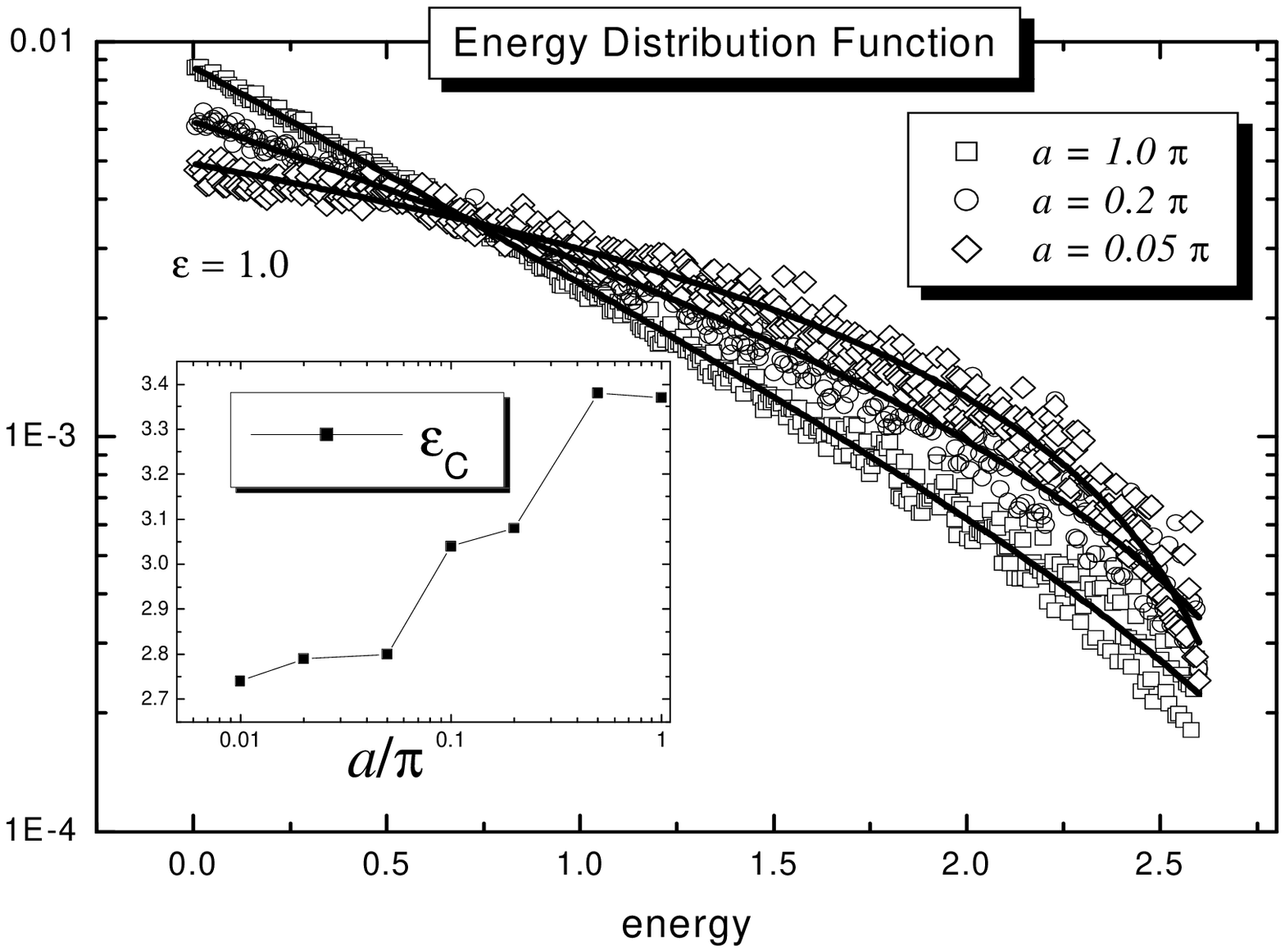}%
\caption{Energy Distribution Function when $\epsilon=1.0$ for different values
of deformation parameter $a$. The insert plot shows the dependence
$\varepsilon_{C}$ \textit{versus} $a$. }%
\label{dcs3michiea}%
\end{center}
\end{figure}

We perform several runs and obtain the energy distribution functions at
different energies, starting from a \textbf{TBIC} initial condition with
$\beta=0.01$. These results are shown in FIG.\ref{dcs3michie}. The reader can
notice that the profile (\ref{michie}) offers a fairly good fitting for all
these dependences, although for the case where $\epsilon=0.4$ some
discrepancies are evident. Such differences probably originate from the method
used to obtain the distribution functions. The insert plot suggests a weak
dependence of the parameter $\varepsilon_{C}\,$\ on the energy per particle
$\epsilon$.

We repeat the calculations by considering different values of the deformation
parameter $a$ and keeping the energy per particle fixed at $\epsilon=1.0$.
These results are shown in FIG.\ref{dcs3michiea}. It is clearly evidenced how
the energy population at large energies is increased when $a$ decreases, which
could be quantitatively described by the decreasing of the parameter
$\varepsilon_{C}$. The origin of this behavior follows from the fact that the
evaporation events take place practically when the involved particles are
those of kinetic energy close to the scape energy $u_{c}$. As already noted,
the effect of the correction to the exponential profile is more noticeable
with the increasing of the energy per particle $\epsilon$. Apparently, the
specific values of $\epsilon$ for the profiles shown in FIG.\ref{dcs2comdfc}
and FIG.\ref{dcs2comdfz} do not allow us to observe this effect, which is more
evident for the results shown in FIG.\ref{dcs3michiea}.

\textit{Summarizing}: \ The above numerical experiments allow us to conclude
that the gas driven by conservative binary encounters and particle evaporation
follows certain quasi-stationary evolution which seems to be macroscopically
controlled by the energy per particle and the particles density. The generic
form of the quasi-stationary profiles is practically independent of the
initial conditions, being appropriately describe by a Boltzmann profile
truncated at the escape energy when the binary encounters favor an effective
exploration of all velocities stable under the evaporation. The reduction of
the effectiveness of this equilibration mechanism produces a continuous
deformation of the truncating Boltzmann profile towards the well-known
Michie-King distribution \cite{bin,king}, which could be qualitatively
described by the profile shown by equation (\ref{michie}).

\section{Generic treatment\label{generic}}

Let us consider the evaporation of the gas of binary encounters in a more
general framework. Let $P_{m}$ be the probability that a given particle be in
the \textit{state} $m$, where $m$ represents certain admissible velocity. The
rate of change of this probability, $\mathcal{K}_{m}\left(  P\right)  $:%

\begin{equation}
\frac{dP_{m}}{d\tau}=\mathcal{K}_{m}\left(  P\right)  ,
\end{equation}
can be divided in two terms as follows:%

\begin{equation}
\mathcal{K}_{m}\left(  P\right)  =\left\langle \text{Eq.Trans}\right\rangle
_{m}+\left\langle \text{NEq.Trans}\right\rangle _{m}. \label{km}%
\end{equation}
The term $\left\langle \text{Eq.Trans}\right\rangle _{m}$ considers all those
binary encounters affecting the state $m$ without provoking evaporation events:%

\begin{equation}
\left\langle \text{Eq.Trans}\right\rangle _{m}=\sum_{AB\tilde{m}}W_{\left(
AB\right)  \left(  m\tilde{m}\right)  }\left(  P_{A}P_{B}-P_{m}P_{\tilde{m}%
}\right)  ,
\end{equation}
while $\left\langle \text{Evap.Trans}\right\rangle _{m}$ involves all those
encounters where a collision particle escapes from the system:%

\begin{equation}
\left\langle \text{Evap.Trans}\right\rangle _{m}=\sum_{ij}D_{\left(
ij\right)  m}P_{i}P_{j}-\sum_{st}D_{\left(  sm\right)  t}P_{s}P_{m}.
\end{equation}

It is very easy to note that the transition probability $W_{\left(  AB\right)
\left(  m\tilde{m}\right)  }$ of the first term involves \textit{four}
admissible states, while the transition probability $D_{\left(  ij\right)  m}$
of second one only involves \textit{three} states. This is the origin of the
particles and energy dissipation by evaporation events, which are described by
the equations:%

\begin{equation}
\frac{dP}{d\tau}=\mathcal{K}\left(  P\right)  ,~\frac{dE}{d\tau}%
=\mathcal{R}\left(  P\right)  \label{papa}%
\end{equation}
being $P=\sum_{m}P_{m}$ and $E=\sum_{m}\varepsilon_{m}P_{m}$, where
$\varepsilon_{m}$ represents the energy of the particle at the \textit{m-th}
state, as well as:%

\begin{equation}
\mathcal{K}\left(  P\right)  =-\sum_{ijm}D_{\left(  ij\right)  m}P_{i}P_{j},
\label{k}%
\end{equation}
and%

\begin{equation}
\mathcal{R}\left(  P\right)  =-\sum_{ijm}D_{\left(  ij\right)  m}P_{i}%
P_{j}\left(  \varepsilon_{i}+\varepsilon_{j}-\varepsilon_{m}\right)  .
\label{p}%
\end{equation}

Let us now introduce the normalized distribution function $f_{m}=P_{m}/P$ (
$\sum_{m}f_{m}=1$ ), which obeys the dynamical equations:%

\begin{equation}
\frac{df_{m}}{d\tau}=P\left[  \mathcal{K}_{m}\left(  f\right)  -f_{m}%
\mathcal{K}\left(  f\right)  \right]  , \label{GTE}%
\end{equation}
where $\mathcal{K}_{m}\left(  f\right)  $ and $\mathcal{K}\left(  f\right)  $
were defined at the equations (\ref{km}) and (\ref{k}).

Two important macroscopic quantities of this system are the energy per
particle $\epsilon\left[  f\right]  $ and the entropy functional $s\left[
f\right]  $:%

\begin{equation}
\epsilon\left[  f\right]  =\sum_{m}\varepsilon_{m}f_{m},~s\left[  f\right]
=-\sum_{m}f_{m}\ln f_{m}.
\end{equation}

\begin{figure}
[t]
\begin{center}
\includegraphics[
height=3.039in,
width=3.2993in
]%
{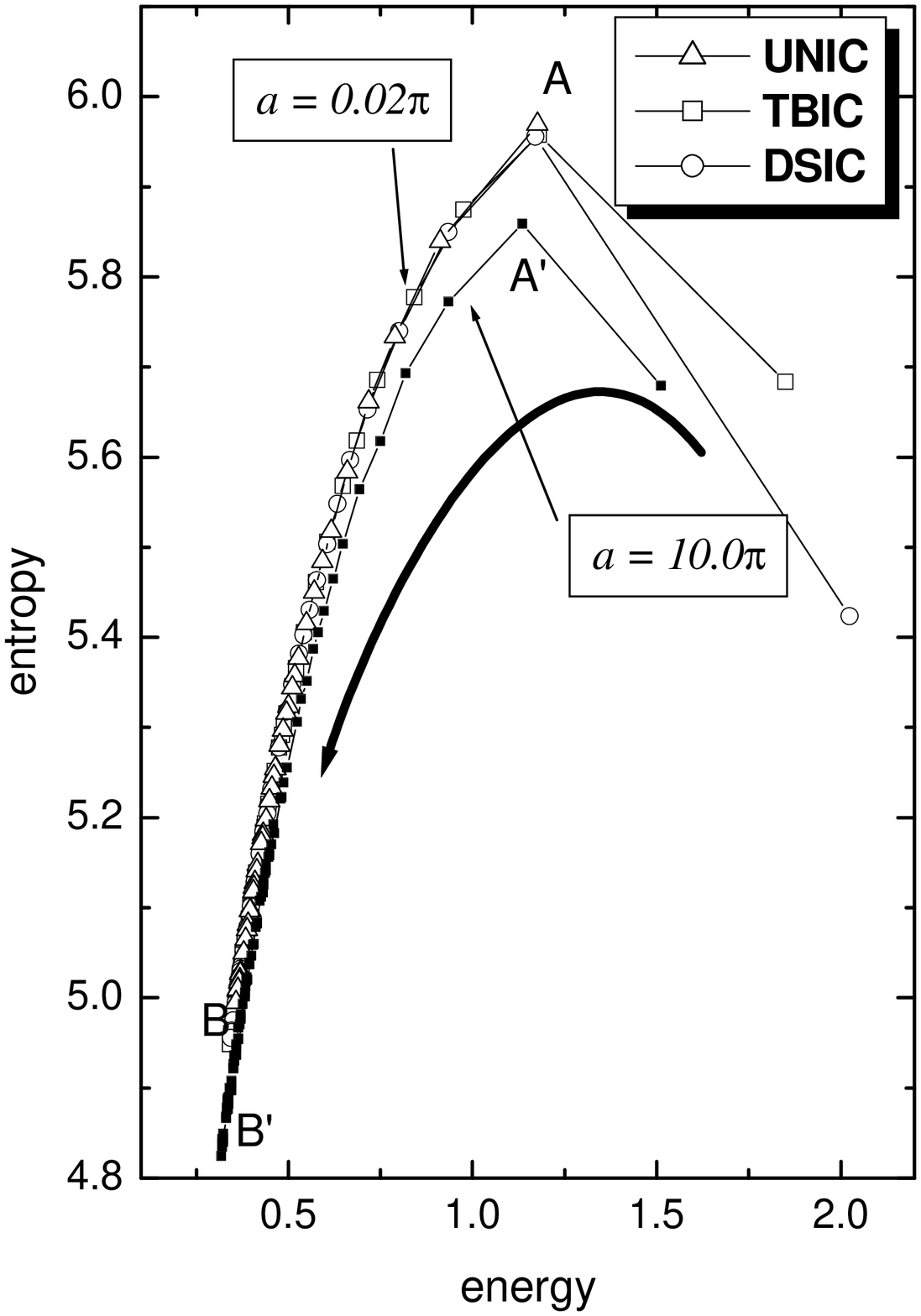}%
\caption{The plane $s$ \textit{versus} $\epsilon$ reveals us that the model
system follows certain sequence \textbf{A-B} along its dynamical evolution
which does not depend on the initial conditions, being this a clear
manifestation of its quasi-stationary character. The curve \textbf{A'-B'} is
the quasi-stationary sequence when $a=10.0\pi$. The thick arrow represents the
direction of the dynamical evolution along these sequences. }%
\label{secuence}%
\end{center}
\end{figure}

Since the distribution function $f_{m}$ must be macroscopically controlled by
the energy per particle $\epsilon$ during the quasi-stationary regime, the
entropy $s\left[  f\right]  $ \textit{versus} $\epsilon$ plane must exhibits
certain sequence revealing a functional dependence among these macroscopic
quantities. FIG.\ref{secuence} confirms this supposition by using three
different initial conditions and considering $a=0.02\pi$ in the differential
cross section (\ref{dCS2}).%

The reader can notice that after a fast relaxation the entropy $s$ evolves
along certain curve \textbf{A-B} in the plane $s$ \textit{versus} $\epsilon$
which does not depend on the initial conditions. This is a clear manifestation
of the quasi-stationary character of the dynamical evolution of the model
system. Despite the reduction of the effectiveness of the equilibration
mechanisms for low values of the deformation parameter $a$, the
quasi-stationary character of the dynamical evolution of the model system
seems to very robust. For comparison, we also show in this figure the
quasi-stationary sequence when $a=10.0\pi$. It is easy to see that the
decreasing of the deformation parameter $a$ provokes an increasing of disorder
for a given energy per particle, which is directly related with the increasing
of the particles population at large energies (see FIG.\ref{dcs3michiea}).

The energy per particle $\epsilon\left[  f\right]  $ evolves according to the equation:%

\begin{equation}
\frac{d}{d\tau}\epsilon\left[  f\right]  =P\left\{  -\mathcal{K}\left[
f\right]  \epsilon\left[  f\right]  +\mathcal{R}\left[  f\right]  \right\}  ,
\end{equation}
being $\mathcal{R}\left[  f\right]  $ the functional of the equation
(\ref{p}), while the dynamics of the entropy functional $s\left[  f\right]  $
is described by the equation:%

\begin{equation}
\frac{d}{d\tau}s\left[  f\right]  =P\left\{  -\mathcal{K}\left[  f\right]
s\left[  f\right]  +\mathcal{W}\left[  f\right]  +\mathcal{D}\left[  f\right]
\right\}  , \label{WD}%
\end{equation}
where $\mathcal{W}\left[  f\right]  $ is given by:%

\begin{align}
\mathcal{W}\left[  f\right]   &  =\frac{1}{4}\sum_{ABm\tilde{m}}W_{\left(
AB\right)  \left(  m\tilde{m}\right)  }\left[  \ln\left(  f_{A}f_{B}\right)
-\ln\left(  f_{m}f_{\tilde{m}}\right)  \right] \nonumber\\
&  \times\left(  f_{A}f_{B}-f_{m}f_{\tilde{m}}\right)  ,
\end{align}
and $\mathcal{D}\left[  f\right]  $ as:%

\begin{equation}
\mathcal{D}\left[  f\right]  =\sum_{\left(  ijm\right)  }D_{\left(  ij\right)
m}f_{i}f_{j}\left(  \ln f_{i}+\ln f_{j}-\ln f_{m}\right)  .
\end{equation}

It is easy to see that the term $\mathcal{W}\left[  f\right]  \geq0$, where
the identity only takes place with the establishment of the \textit{detailed
balance}:%

\begin{equation}
f\left(  \mathbf{v}_{1}\right)  f\left(  \mathbf{v}_{2}\right)  =f\left(
\mathbf{\tilde{v}}_{1}\right)  f\left(  \mathbf{\tilde{v}}_{2}\right)  ,
\end{equation}
among all those pairs of velocities $\left(  \mathbf{v}_{1},\mathbf{v}%
_{2}\right)  $ and $\left(  \mathbf{\tilde{v}}_{1},\mathbf{\tilde{v}}%
_{2}\right)  $ belonging to the admissible space $\mathbf{\Sigma}$ which are
related by a collision event without evaporation. On the other hand, the
functional $\mathcal{D}\left[  f\right]  $ does not exhibit, generally
speaking, a definite signature. The energies $\varepsilon_{m}$,~$\varepsilon
_{i}$ and $\varepsilon_{j}$ corresponding to the states related by an
evaporation event, $\left(  ij\rightarrow m\right)  $ are ordered as follows:
$\varepsilon_{m}<\min\left(  \varepsilon_{i},\varepsilon_{j}\right)  $. It is
not difficult to show that $\mathcal{D}\left[  f\right]  \leq0$ (or
$\mathcal{D}\left[  f\right]  \geq0)$~whenever $f\left(  \varepsilon\right)  $
be a decreasing (or increasing) monotonic function on $\varepsilon$,
requirement satisfied by the Michie-King-like profile (\ref{michie}).

The quasi-stationary regime is established by the competition of\textit{ two
tendencies}: First: the evolution towards the most likely macroscopic
configuration where the detailed balance is imposed, which is represented by
the presence of the conservative term $\mathcal{W}\left[  f\right]  $ in the
equation (\ref{WD}); Second one: the tendency of finding more stability under
the evaporation, related with the presence of the dissipative terms
$-\mathcal{K}\left[  f\right]  s\left[  f\right]  +$ $\mathcal{D}\left[
f\right]  $. The functionals $\mathcal{K}\left[  f\right]  $, $\mathcal{D}%
\left[  f\right]  $ and $\mathcal{R}\left[  f\right]  $, despite being
independent, vanish simultaneously when a monotonic distribution function on
the energy $f\left(  \varepsilon\right)  $, which drops to zero in the range
$u_{c}/2\leq\varepsilon\leq u_{c}$, is considered. Since the stability of
these tendencies cannot be simultaneously satisfied, the system evolves
towards certain intermediate distribution which is never stationary.

It is well-known the difficulties to find an appropriate criterion leading to
the theoretical prediction of quasi-stationary profiles. We only found an
appropriate criterion when the system equilibration mechanism is effective in
exploring all accessible configurations which are stable under the
evaporation: the system practically reach the most probable one, being this
configuration the quasi-stationary profile. In those cases, the contribution
of the dissipative terms in the entropy dynamical evolution seems to be only
comparable to the term $\mathcal{W}\left[  f\right]  $ for configurations
close to the most likely because of the predominance of binary transitions
which does not involve particle evaporation. However, the picture turns out to
be considerably difficult in the general case with the reduction of the
effectiveness of the equilibration mechanisms.

As already stressed, during the dynamical evolution the system distribution
functions will converge towards certain sequence which is identified with the
quasi-stationary one. Once settled at a given point of this sequence, the
system will evolve along it. If we know some configuration belonging to this
sequence, the progressive evolution of the system will reproduce the rest of
the quasi-stationary profiles. It is not difficult to understand that this
quasi-stationary sequence \textit{must contain} the microscopic configuration
with the maximum energy per particle $\epsilon$, $f_{m}\left(  e\right)  $.
Since $0\leq\epsilon\leq u_{c}$, the distribution function with maxima energy
per particle is $f_{m}\left(  e\right)  =\delta\left(  e-u_{c}\right)  $.
Thus, in our opinion, the problem of finding the quasi-stationary profiles can
be reduced to the determination of the temporal dependence of the distribution
function $f\left(  e;\tau\right)  $ by using the dynamical equation
(\ref{GTE}) with $f\left(  e;0\right)  =\delta\left(  e-u_{c}\right)  $, but
this problem does not have a general analytical solution because of the
nonlinear nature of the dynamics.%

\begin{figure}
[t]
\begin{center}
\includegraphics[
height=3.039in,
width=3.2958in
]%
{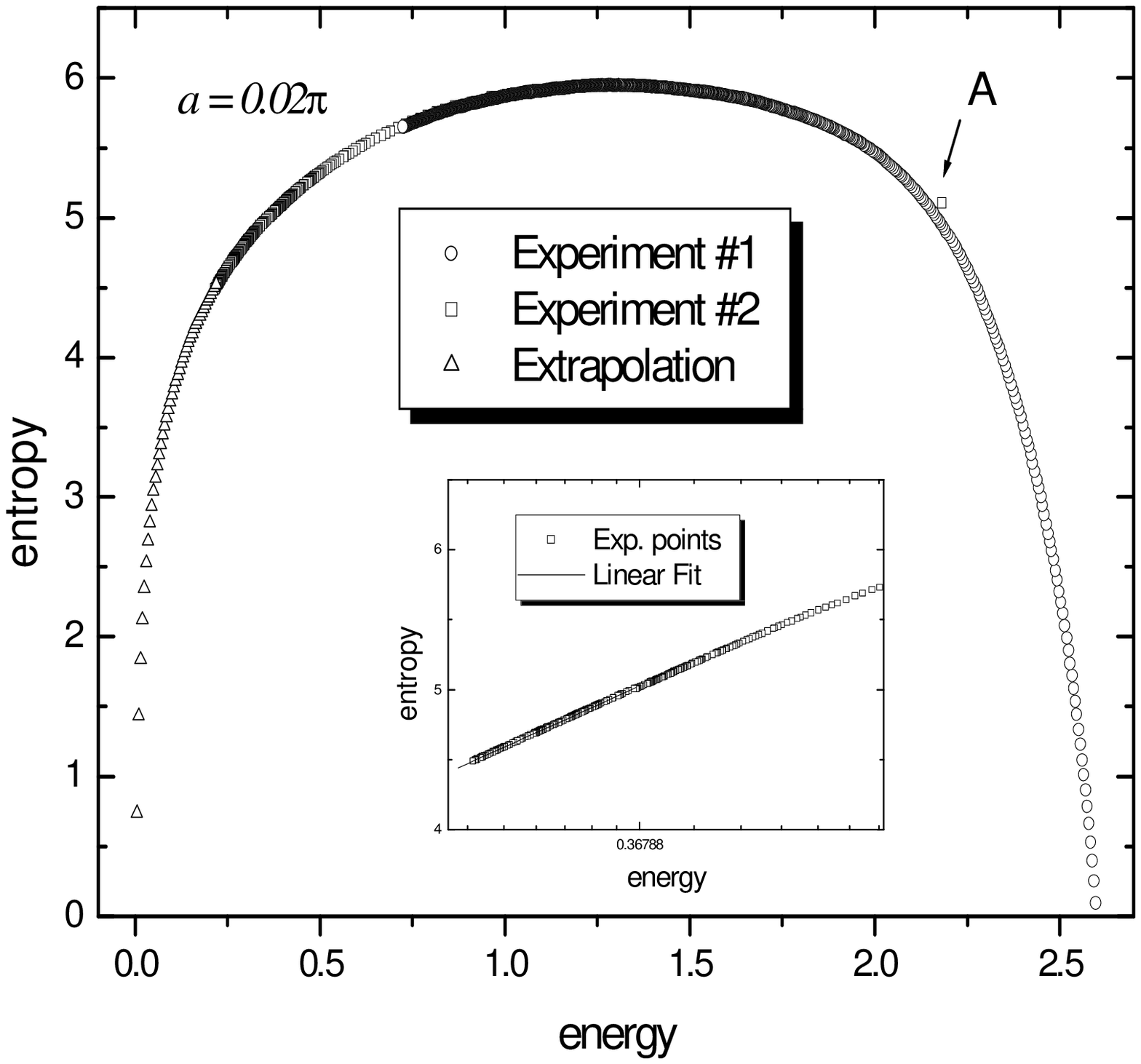}%
\caption{$s_{qs}$ \textit{versus} $\epsilon$ dependence for the
quasi-stationary evolution obtained by using the dynamical criteria when
$a=0.02\pi$. The point $A$ corresponds to the starting configuration for the
second experiment. The insert plot shows the linear dependence at low energies
for the $s$ \textit{versus} $\ln\epsilon$.}%
\label{entropy}%
\end{center}
\end{figure}

We use the above dynamical criterion to build what we consider as the
quasi-stationary sequence $s_{qs}$ \textit{versus} $\epsilon$, which is shown
in FIG.\ref{entropy} by considering $a=0.02\pi$. We obtain the data from two
dynamical experiments, as well as by the extrapolation to the low energies
region. The first experiment was performed by considering $\mu=0.005$ in order
to reduce the rhythm of the dynamical evolution when $\epsilon$ is close to
$u_{c}$, and the second one, by considering $\mu=0.1$. The reader can recall
that this parameter controls the temporal increase for each computational
step, but it does not affect the dependence of distribution function on the
energy per particle.

Both experiments were performed by considering $N=20000$ and by dividing the
range $0\leq e\leq u_{c}$ in $400$ intervals in order to build the
distribution functions. The function $f_{m}\left(  e\right)  $ was
approximated by using a \textbf{DSIC} with $E_{2}=u_{C}$ and $E_{1}%
=u_{c}-u_{c}/400$. The second experiment starts from a \textbf{TBIC} with
$\beta=-10.0$, which\ exhibits a fast convergence towards the quasi-stationary
sequence of the first experiment. Notice that in this figure there is a
superposition of both experiments in the intermediate energetic range and no
discrepancy is evidenced. The range $0<\epsilon\leq0.22$ was obtained from the
extrapolation of the experimental data because of the evolution is extremely
slow in this interval, and basically, the distribution functions do not differ
significantly from the truncating Boltzmann profile, yielding $s\simeq
6.035+\ln\epsilon$ in this energetic range.

\section{Nonhomogeneous character of the confining
potential\label{nonhomogeneous}}

The model system considered in the above sections disregards the effect of the
nonhomogeneous character of the confining potential energy. We are interested
in describing this effect, at least, in the extreme case where the system
equilibration mechanism is very effective. As already observed in the
numerical experiments, when the particles are able to explore in an effective
fashion all those microscopic configurations which are stable under the
evaporation, the progressive evolution of the system leads to the setting of
the ordinary equilibrium conditions, as an example, the imposition of the
detailed balance in the framework of the binary encounters.

In the context of Hamiltonian systems, the equilibration mechanisms relay on
the strong chaoticity of the microscopic dynamics. This property usually leads
to the ergodicity and the equilibrium conditions are ordinarily described by
considering a microcanonical distribution \cite{pet}. If evaporation is
present, the above picture can be naturally generalized by considering a
microcanonical distribution where it has been disregarded all those
microscopic configurations where the particles are able to escape. The above
argument was used in a very recent paper \cite{alt} in the context of
astrophysical systems in order to develop an alternative version of the
isothermal model of Antonov \cite{antonov}.

The stable configurations for a Hamiltonian model of the form:%

\begin{equation}
H_{N}=K_{N}+V_{N}=\sum_{k=1}^{N}\frac{1}{2m}\mathbf{p}_{k}^{2}+\sum
_{i=1}^{N-1}\sum_{j>i}^{N}\phi\left(  \mathbf{r}_{j}-\mathbf{r}_{i}\right)  ,
\end{equation}
is the subset of the N-body phase space where the particles satisfy the
inequalities: $\mathbf{p}_{k}^{2}/2+u\left(  \mathbf{r}_{k}\right)
<\epsilon_{S}$, being $\epsilon_{S}$ the energy cutoff, $\mathbf{p}_{k}$ and
$u\left(  \mathbf{r}_{k}\right)  $, the linear momentum and potential energy
for the \textit{k-th} particle:
\begin{equation}
u\left(  \mathbf{r}_{k}\right)  =\sum_{i\not =k}\phi\left(  \mathbf{r}%
_{k}-\mathbf{r}_{i}\right)  ,
\end{equation}
where $k=1,...N$. We are interested in the specific form of the
quasi-stationary distribution function $f\left(  \mathbf{r},\mathbf{p}\right)
$ in such conditions.

This question was solved in ref.\cite{alt} in the context of astrophysical
systems by considering the mean field approximation appearing when $N$ tends
to infinity. The generalization of these results is straightforwardly followed
by taking into account the generalized form of the mean field potential energy
$u\left(  \mathbf{r}\right)  $:%

\begin{equation}
u\left(  \mathbf{r}\right)  =\int d^{3}\mathbf{x}~\phi\left(  \mathbf{r-x}%
\right)  \rho\left(  \mathbf{x}\right)  ,
\end{equation}
where $\rho\left(  \mathbf{r}\right)  $ is the particles density at the point
$\mathbf{r}$ of the physical space. Therefore, we will only expose here the
final results. The interested reader can see the ref.\cite{alt} for details of
this analysis in the framework of the Newtonian interaction.

It can be proved that the particles density $\rho\left(  \mathbf{r}\right)  $
is given by the expression:%

\begin{equation}
\rho\left(  \mathbf{r}\right)  =\left(  \frac{m}{2\pi\hbar^{2}\beta}\right)
^{\frac{3}{2}}\exp\langle-\mu+C\left(  \mathbf{r}\right)  -\beta u\left(
\mathbf{r}\right)  \rangle F\left[  \sqrt{\Phi\left(  \mathbf{r}\right)
}\right]  , \label{ro}%
\end{equation}
where $\beta$ and $\mu$ are the canonical parameters which are specified by
the energy and particles number constraints, $F\left[  z\right]  $ is defined
by the integral:%

\begin{equation}
F\left[  z\right]  =\pi^{-\frac{3}{2}}\int_{0}^{z}4\pi x^{2}\exp\left(
-x^{2}\right)  dx,
\end{equation}
and $\Phi\left(  \mathbf{r}\right)  =\beta\left[  \epsilon_{S}-u\left(
\mathbf{r}\right)  \right]  $. The function $C\left(  \mathbf{r}\right)  $ is
obtained in a \textit{self-consistent} way as follows:%

\begin{equation}
C\left(  \mathbf{r}\right)  =\int d^{3}\mathbf{x}~\phi\left(  \mathbf{r-x}%
\right)  \rho\left(  \mathbf{x}\right)  \frac{\partial}{\partial u\left(
\mathbf{x}\right)  }\ln F\left[  \sqrt{\Phi\left(  \mathbf{x}\right)
}\right]  .
\end{equation}

Since $\rho\left(  \mathbf{r}\right)  =\int d^{3}\mathbf{p}~f\left(
\mathbf{r},\mathbf{p}\right)  $, it is very easy to note that equation
(\ref{ro}) is derived from the distribution function $f\left(  \mathbf{r}%
,\mathbf{p}\right)  $:%

\begin{equation}
f\left(  \mathbf{r},\mathbf{p}\right)  = C_{0}\exp\left(  C\left[
\mathbf{r}\right)  - \beta\epsilon\left(  \mathbf{r},\mathbf{p}\right)
\right]  . \label{fcc}%
\end{equation}
which vanishes when $\epsilon\left(  \mathbf{r},\mathbf{p}\right)
>\epsilon_{S}$, being $\epsilon\left(  \mathbf{r},\mathbf{p}\right)  =\frac
{1}{2m}\mathbf{p}^{2}+u\left(  \mathbf{r}\right)  $. This expression differs
from the truncating isothermal profile only because of the presence of the
self-consistent function $C\left(  \mathbf{r}\right)  $, which appears as an
additional consequence of the particle evaporation due to the nonhomogeneous
character of the global interaction. Although the $\rho$ \textit{versus}
$u\left(  r\right)  $ dependence (\ref{ro}) unifies the isothermal with the
polytropic dependencies: $\rho\sim\exp\left(  C+\Phi\right)  $ when $\Phi$ is
large enough, while $\rho\sim\exp\left(  C\right)  \Phi^{\frac{3}{2}}$ when
$\Phi<<1$, the presence of the function $C$ introduces a sensitive
modification to the features of the solutions.

\begin{figure}
[t]
\begin{center}
\includegraphics[
height=3.2396in,
width=3.2396in
]%
{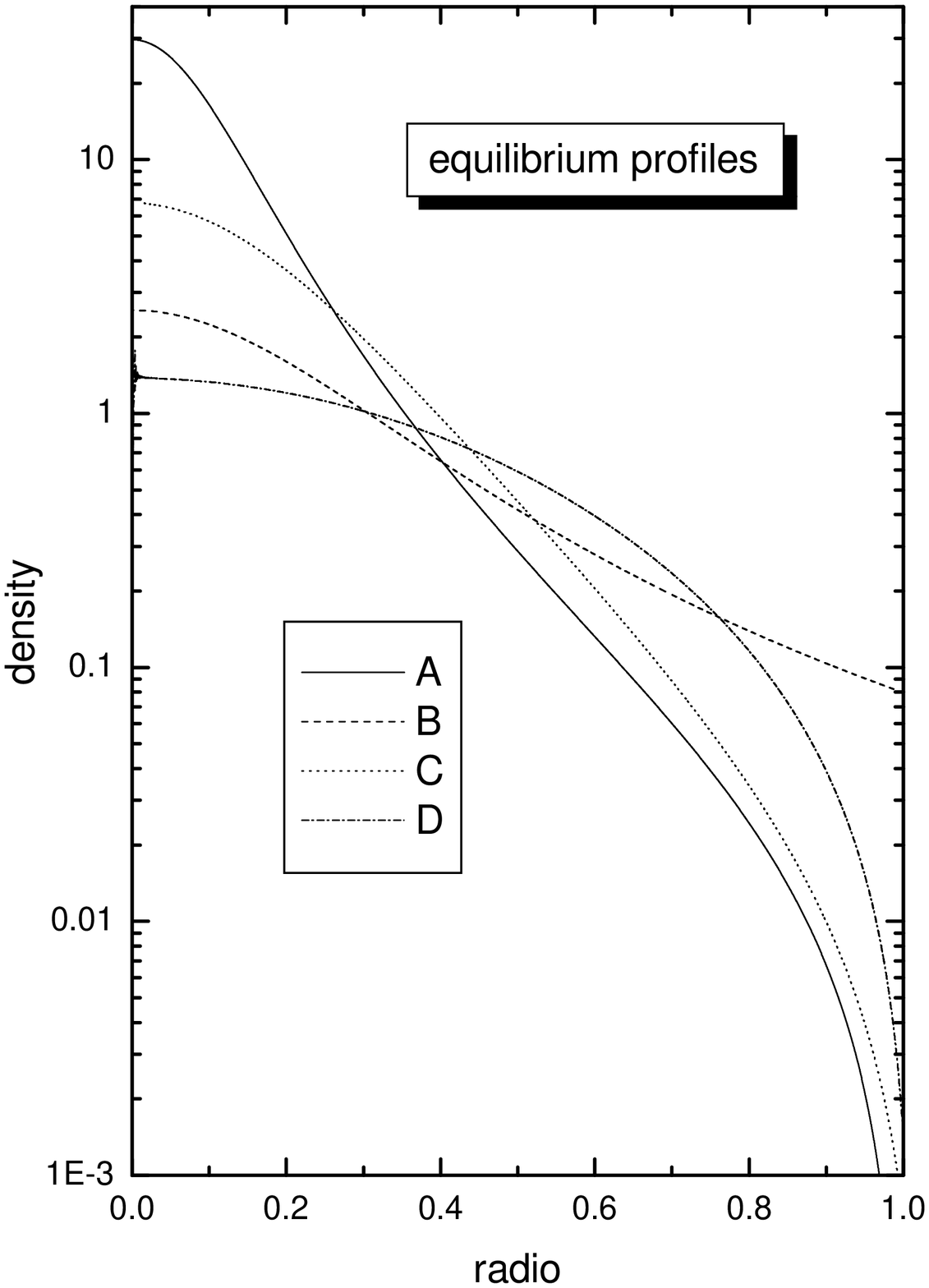}%
\caption{The effect of the function $C\left(  \mathbf{r}\right)  $ in the
particles density for the Newtonian potential.}%
\label{com_prof}%
\end{center}
\end{figure}

FIG.\ref{com_prof} shows the effect of the function $C\left(  \mathbf{r}%
\right)  $ in the particles density for the Newtonian potential $\phi\left(
r\right)  =-\kappa/r$ for configurations with spherical symmetry (details of
this study are also found in ref.\cite{alt}). The profiles A and C were
obtained by considering the distribution function (\ref{fcc}): profile A
corresponds to a quasi-stationary configuration with low energy, which is
characterized by the existence of an isothermal core and a polytropic halo,
while C corresponds to a high energy configuration where the isothermal core
is not present. Profiles B and D correspond respectively to an isothermal and
polytropic configurations with polytropic index $\gamma=\frac{5}{3}$
\ \cite{bin} (characterized by the dependence $\rho\left(  \Phi\right)
\propto\Phi^{\frac{3}{2}}$). The presence of the function $C\left(
\mathbf{r}\right)  $ leads to a significant concentration of the particles
toward the inner regions which goes beyond what the isothermal or the
polytropic models predict.%

It is easy to understand that the function $C$ depends in a self-consistent
way on the energy function $\epsilon\left(  \mathbf{r},\mathbf{p}\right)  $.
Therefore, the distribution function $f\left(  \mathbf{r},\mathbf{p}\right)
$, as a whole, depends only on the function $\epsilon\left(  \mathbf{r}%
,\mathbf{p}\right)  $, that is, $f\left(  \mathbf{r},\mathbf{p}\right)
=\mathcal{F}\left[  \epsilon\left(  \mathbf{r},\mathbf{p}\right)  \right]  $.
Needless to say that this is a requirement of the well-known Jeans
theorem\ \cite{jeans}. However, the consideration of such effect only starting
from this last theorem is very unlikely. The present analysis suggests that
the simple generalization of the homogeneous distribution functions $f\left(
\mathbf{r},\mathbf{p}\right)  =\varphi\left(  \frac{1}{2m}\mathbf{p}%
^{2}\right)  $, by substituting $\frac{1}{2m}\mathbf{p}^{2}\rightarrow\frac
{1}{2m}\mathbf{p}^{2}+u\left(  \mathbf{r}\right)  $ may be extremely naive,
and probably, most of these simple extensions are disregarding important
effects of the nonhomogeneous character of the global interaction in presence
of particle evaporation.

Thus, the consideration of a weaker chaoticity\ of the microscopic dynamics
and the nonhomogeneous character of the interactions involve serious
difficulties in the theoretical prediction of the quasi-stationary profiles.
The reader may notice that the approach developed in ref.\cite{alt} and the
standard treatment of some astrophysical structures, such as globular clusters
and elliptical galaxies, by using the Michie-King profile \cite{king,bin}, are
extreme alternatives to consider the effect of the evaporation in such a
framework. The applicability of such models depends crucially on how effective
is the equilibration mechanism in an actual astrophysical systems.

It is well-established that the picture of the binary encounters provide a
nice justification for the use of the King's models. The basic explanation is
found in the fact that the long-range character of the gravitational
interaction favors much more the small energy interchange among those distant
particles in comparison to the large energy interchanges occurring during the
close encounters \cite{spitzer,king,chandra}.

Nevertheless, the binary encounter is not the only equilibration mechanism
appearing in this context. In fact, it has been also shown the existence of a
very strong chaoticity in gravitational N-body systems \cite{pet}, which acts
in time scales much smaller than the characteristic time for the close binary
collisions, as well as it operates for every bound energy. In the viewpoint of
the author of such work, this result not only confirms that "... \textit{bound
collisionless selfgravitating systems are characterized by a dynamical
instability proceeding at a very fast rate, but also suggests that they
possess the strongest statistical properties, in analogy with those of
standard dynamical systems in the regime of fully developed stochasticity ..."
}\cite{pet}. As expected, this argument supports the consideration of the
ergodic hypothesis in this context, and therefore, the consideration of a
suitably regularized microcanonical description for actual astrophysical
systems \cite{pet,gross,de vega,chava}.

Despite the differences concerning to the specific form of the above energy
profiles, the question is that the features of their corresponding spatial
particle distributions possess more similarities than differences: both
distributions lead to finite configurations with isothermal cores and
polytropic haloes, even purely polytropic configurations are\ also possible,
where the system size is determined from the tidals interactions.

In the remarkable paper \cite{king}, King performed a numerical comparison of
the Michie-King model with the isothermal distribution truncated at the escape
energy, without the consideration of self-consistent function $C$, which is
shown in FIG.2 and FIG.7 of this reference. While the King model provides an
excellent fit for the data corresponding to a high concentrate globular
cluster, the truncated distribution is unable to fit an observational data
where a larger concentration of particles appears towards the inner region. In
our view, this could be the very effect introduced by taking into
consideration the self-consistent function $C$, but this last observation
deserves a further analysis. Moreover, it is difficult to distinguish the
origin of such effect since there exist other physical backgrounds modifying
the form of the particles distribution in the space, such as the mass
segregation \cite{grif} (see also ref.\cite{jordan} for a recent development).

\section{Conclusions}

The numerical experiments carried out for the gas of binary encounters clearly
show the robustness of the quasi-stationary character of the dynamical
evolution of this model system. The specific form of the quasi-stationary
distribution functions will depend crucially on the effectiveness of the
microscopic dynamics in exploring all those microscopic configurations which
are stable under the particle evaporation.

An efficient microscopic dynamics leads to the imposition of the detailed
balance for all those binary transitions which do not involve particle
evaporation, while a decreasing of this effectivity leads to quasi-equilibrium
profiles whose specific form depends on the details of the dynamical equations.

The analysis, developed in section \ref{nonhomogeneous}, in order to take into
account the effect of the nonhomogeneous character of the confining
interactions for those Hamiltonian systems driven by a strong chaotic
dynamics, suggests the appearance of nontrivial effects of the evaporation in
the spatial particles distribution. The contribution of such nontrivial
effects cause a significant modification of the quasi-stationary profiles, as
already shown in the context of the gravitational interaction.

The above results show how difficult could be the theoretical study of the
evaporation effect in a general physical background, mainly, when the
equilibration mechanisms are not so effective and the dynamics is driven by
the presence of long-range forces.

\begin{acknowledgments}
L. Velazquez is grateful for the hospitality of the \textit{Group of
Nonextensive Statistical Mechanics} head by C. Tsallis during his visit to the
CBPF. He also acknowledges the ICTP/CLAF financial support. HJMC thanks FAPERJ
(Brazil) for a Grant-in-Aid.
\end{acknowledgments}

\end{document}